\documentclass{article}

	\usepackage{amsmath}
	\usepackage{epsfig}
	\usepackage{color}
	\usepackage{float}
	\usepackage[table]{xcolor}
	\usepackage[margin=1.5in]{geometry}
	\usepackage[utf8]{inputenc}
	\usepackage{siunitx}
	\usepackage{textgreek}
	\usepackage{upgreek}
	\usepackage{graphicx}
	\usepackage{subcaption}
	\usepackage[toc,page]{appendix}
	\usepackage[affil-it]{authblk}
	\usepackage[english]{babel}
	\usepackage{blindtext}
	\newcommand{\bd}{~\mathbf{d}}
	\newcommand{\bRp}{\mathbf{R}^+}
	\newcommand{\bRn}{\mathbf{R}^-}
	\newcommand{\bR}{\mathbf{R}}
	\newcommand{\brp}{\mathbf{r}^+}
	\newcommand{\brn}{\mathbf{r}^-}  
	\newcommand{\br}{\mathbf{r}}
	\newcommand{\be}{~\mathbf{e}}
	\newcommand{\pardev}[2]{\frac{\partial{#1}}{\partial{#2}}}
	\newcommand{\bP}{\mathbf{P}}
	\newcommand{\bQ}{\mathbf{Q}}
	\newcommand{\bZ}{\mathbf{Z}}
	\newcommand{\bI}{\mathbf{1}}

	\newcommand{\bin}{\boldsymbol{n}}
	\newcommand{\binn}{\boldsymbol{n}^-}
	\newcommand{\binp}{\boldsymbol{n}^+}
	\newcommand{\binc}{\boldsymbol{n}^c}

	\newcommand{\bim}{\boldsymbol{m}}
	\newcommand{\bimn}{\boldsymbol{m}^-}
	\newcommand{\bimp}{\boldsymbol{m}^+}
	\newcommand{\bimc}{\boldsymbol{m}^c}
	
	\newcommand{\biw}{\boldsymbol{w}}
	\newcommand{\bw}{\mathbf{w}}

	\newcommand{\bc}{\boldsymbol{c}}
	\newcommand{\bfo}{\boldsymbol{f}}
	
	\newcommand{\bfv}{~\mathbf{f}}

	\newcommand{\cosr}{\cos \omega x}
	\newcommand{\sinr}{\sin \omega x}

	\newcommand{\bfone}{(\cosr\bd_1+\sinr\bd_2)}
	\newcommand{\bftwo}{(-\sinr\bd_1+\cosr\bd_2)}
	
	\newcommand{\er}{(\cosr\be_1+\sinr\be_2)}
	\newcommand{\ethe}{(-\sinr\be_1+\cosr\be_2)}

	\newcommand{\tp}{\boldsymbol{t}^+}
	\newcommand{\np}{\boldsymbol{n}^+}
	\newcommand{\bp}{\boldsymbol{b}^+}
	
	\newcommand{\tpz}{\boldsymbol{t}^+_0}
	\newcommand{\npz}{\boldsymbol{n}^+_0}
	\newcommand{\bpz}{\boldsymbol{b}^+_0}

	\newcommand{\tn}{\boldsymbol{t}^-}
	\newcommand{\nn}{\boldsymbol{n}^-}
	\newcommand{\bn}{\boldsymbol{b}^-}
	
	\newcommand{\tnz}{\boldsymbol{t}^-_0}
	\newcommand{\nnz}{\boldsymbol{n}^-_0}
	\newcommand{\bnz}{\boldsymbol{b}^-_0}

	\newcommand{\oeps}{O(\varepsilon)}
	\newcommand{\oepss}{O(\varepsilon^2)}
	
	\newcommand{\thetap}{\mathbf{\Theta^+}}
	\newcommand{\thetan}{\mathbf{\Theta^-}}
	\newcommand{\bphic}{\mathbf{\Phi^c}}
	\newcommand{\bphi}{\mathbf{\Phi}}
	\newcommand{\phic}{\pmb{\phi^c}}
	\newcommand{\etabar}{\pmb{\bar{\eta}}}
	\newcommand{\boeta}{\pmb{\eta}}
	
	\newcommand{\betap}{\beta^+}
	
	\newcommand{\negs}{^-}
	\newcommand{\poss}{^+}

	\newcommand{\bv}{\mathbf{v}}

	\newcommand{\sinabt}{\sin \frac{\alpha}{2}}
	\newcommand{\cosabt}{\cos \frac{\alpha}{2}}
	\newcommand{\sinrabt}{\sin(\omega x+ \frac{\alpha}{2})}
	\newcommand{\cosrabt}{\cos(\omega x+ \frac{\alpha}{2})}
	\newcommand{\bfonet}{\Big(\sin(\omega x+\frac{\alpha}{2})\bd_1-\cos(\omega x+\frac{\alpha}{2})\bd_2\Big)}
	\newcommand{\bftwot}{\Big(\cos(\omega x+\frac{\alpha}{2})\bd_1+\sin(\omega x+\frac{\alpha}{2})\bd_2\Big)}
	
	\newcommand{\eqsp}[1]{\begin{equation}
		\begin{split}
		#1
		\end{split}
		\end{equation}}
	\newcommand{\nm}{\text{ nm}}

	\newcommand{\biv}{\boldsymbol{v}}

	\newcommand{\bU}{\mathbf{U}}
	
	\newcommand{\bet}{\pmb{\upeta}}
	\newcommand{\biet}{\pmb{\eta}}
	
	\newcommand{\bhk}{\hat{\mathbf{k}}}

		\title{Allosteric interactions in a birod model of DNA}
		
		\author[1]{Jaspreet Singh}
		\author[1,*]{Prashant K. Purohit}
		
		\affil[1]{Department of Mechanical Engineering \& Applied Mechanics,
			University of Pennsylvania, Philadelphia, PA 19104.}
		\affil[*]{Corresponding author: Prashant K. Purohit, purohit@seas.upenn.edu}
		
\begin{document}
			\maketitle
		\begin{abstract}
			Allosteric interactions between molecules bound to DNA at distant locations have been known for a long time. The phenomenon has been studied via experiments and numerical simulations, but a comprehensive understanding grounded in a theory of DNA elasticity remains a challenge. Here we quantify allosteric interactions between two entities bound to DNA by using the theory of birods. We recognize that molecules bound to DNA cause local deformations that can be captured in a birod model which consists of two elastic strands interacting via an elastic web representing the base-pairs. We show that the displacement field caused by bound entities decays exponentially with distance from the binding site. We compute the interaction energy between two proteins on DNA as a function of distance between them and find that it decays exponentially while oscillating with the periodicity of the double-helix, in excellent agreement with experiments. The decay length of the interaction energy can be determined in terms of the mechanical properties of the strands and the webbing in our birod model, and it varies with the GC content of the DNA. Our model provides a framework for viewing allosteric interactions in DNA within the ambit of configurational forces of continuum elasticity.  
		\end{abstract}
		%%%%%%%%%%%%%%%%%%%%%%%%%%%
		
		%%%%%%%%%% Insert the texts which can accomdate on firstpage in the tag "fmtext" %%%%%

		%%%%%%%%%%%%%%% End of first page %%%%%%%%%%%%%%%%%%%%%

		\section{Introduction}
		Configurational forces that describe the interaction between defects in an elastic solid 
		are those that depend explicitly on the positions of the defects~\cite{phillipsbook,gurtin}. For example, two parellel screw
		dislocations at a distance $a$ from each other interact with a configurational force per unit length 
		proportional to $1/a$ or an energy per unit length proportional to $\log a$~\cite{weertman}. Similarly, 
		the interaction energy of a point defect located at distance $a$ from an edge dislocation varies as
		$1/a$. Just as defects produce local elastic fields in a solid,
		proteins binding to DNA also deform it locally. Since DNA behaves like an elastic rod 
		at scales of a few tens of nanometers \cite{philnel}, we expect
		that if two proteins bind to DNA separated by a distance $a$ then the
		deformation fields created by them will overlap and lead to an interaction energy which
		depends on $a$ in a clearly quantifiable way. This problem has not been theoretically 
		addressed so far, but there is experimental evidence of the interaction. Some of this 
		experimental evidence has been extracted by connecting the interaction energy with the 
		kinetics of protein binding/unbinding. In spirit, this is similar to continuum 
		elasticity in which configurational forces often determine defect dynamics 
		through a kinetic law~\cite{phillipsbook,gurtin}. Kim {\it et al.}~\cite{sunscience} have exploited
		this connection of interaction energies to kinetics to show that gene expression, which 
		depends on RNA polymerase binding affinity to DNA
		in live bacteria, is a function of the proximity of LacR and T7 RNA polymerase bound to DNA. 
		Similarly, the IHF protein affects RNA polymerase activity in {\it E. Coli} DNA~\cite{ihf}. 
		Again, the binding of a drug distamycin to calf thymus DNA has been 
		shown to be cooperative i.e., if one drug molecule binds to the DNA, then it becomes energetically
		favorable for other drug molecules to bind~\cite{crothers1979}. Similarly, binding of the 
		Hox transcription factor to DNA contributes nearly 1.5 kcal/mol to binding the Exd transcription 
		factor~\cite{hoxexd}. These effects are called {\it allosteric} interactions on DNA. 
		Our goal in this paper is to quantify interaction energies between proteins binding to DNA
		as a function of the distance $a$ separating them and the boundary conditions imposed by 
		the proteins on the DNA. We will apply our methods to the quantitative experimental results 
		of Kim {\it et al.}\cite{sunscience} who measure allosteric effects on gene expression as well as 
		transcription factor affinity to DNA.
		
		In the experiments of Kim {\it et al.}~\cite{sunscience} one end 
		of a DNA molecule is attached to the passivated surface of a flow cell and binding sites are 
		provided for two specific proteins to bind. The length
		of the DNA between these binding sites, $a$, is increased in 1bp increments between 7 base-pairs (bp) 
		and 45bp. First, one type of fluorescently labeled protein (call it A) is flowed into the cell so 
		that it binds to the DNA. Then, the second protein (call it B) is flowed in at a specific 
		concentration. The dissociation times of the fluorescent protein are then monitored as a 
		function of $a$. This dissociation time depends on the free-energy change $\Delta G$ of the 
		DNA + two protein complex from the state when the two proteins are bound to that when 
		protein A is unbound. Now, in general, the free energy $\Delta G$ of the ternary complex 
		formed by the DNA and proteins A and B consists of three parts~\cite{sunscience}:
		\begin{equation}
		\Delta G = \Delta G_A + \Delta G_B + \Delta\Delta G_{AB}(a),
		\end{equation}    
		where $\Delta G_{A}$ and $\Delta G_{B}$ are the free energy changes caused by binding of A and
		B respectively to the DNA. These are constants. The last term $\Delta\Delta G_{AB}(a)$ is the portion
		of the free energy change that accounts for the interaction of the two proteins bound to the DNA
		while being separated by a distance $a$. The off-rate of A, which is affected by this term, is plotted as a 
		function of $a$ in Kim {\it et al.}~\cite{sunscience} and it is found that it oscillates with a period
		of 10-11bp with the amplitude of oscillation decreasing as a function of $a$. Similar curves for a free
		energy as a function of separation between protein binding sites on DNA have been obtained experimentally
		for the binding of the {\it lac} repressor to DNA\cite{muller_lacR_expt,maher_expt}. It has been shown that these 
		free energy profiles can be reproduced by modeling DNA as an elastic rod which is forced into forming
		a loop due to stereo-specific binding of the ${\it lac}$ repressor monomers \cite{purohit_nelson} which come
		together due to thermal fluctuations. However,
		Kim {\it et al.}~\cite{sunscience} have ruled out DNA loop formation by careful experimental design and choice of 
		DNA binding proteins. They have also found that the form of the curve is independent 
		of ionic strength (ruling out electrostatic interactions between A and B), but dependent on modifications of the linker DNA. 
		Kim {\it et al.}~\cite{sunscience} infer that this implies $\Delta\Delta G_{AB}(a)$ largely depends 
		on DNA mechanical properties. However, as yet there is no analytical description of how the interaction
		energy $\Delta\Delta G_{AB}(a)$ depends on the DNA mechanical properties.  
		
		Allosteric effects and their relation to protein DNA interactions have been studied using molecular dynamic (MD) 
		simulations \cite{jpc_simulations,nacid_simulations}. Gu \textit{et al.}\cite{jpc_simulations} have studied 
		various kinds of deformations which include shift, roll, rise, twist, slide, and tilt of the DNA bases. 
		They observed a sinusoidal 
		correlation in the major groove widths similar to the one observed by Kim \textit{et al.}\cite{sunscience}. Furthermore, 
		Gu \textit{et al.} point out that the presence of GC rich sequences dampens the allosteric effects which 
		is what Kim \textit{et al.} observe experimentally. Major groove widths have also been implicated in the 
		MD simulations of Hancock \textit{et al.}~\cite{nacid_simulations} who show how bound proteins alter this quantity.
		In contrast, our approach in this paper is based on elastic energy considerations and could compliment the
		analysis of major groove widths as an indicator of allostery in DNA. 
		
		Our goal in this paper is to quantitatively describe allosteric interactions using 
		the {\it birod} model of DNA of Moakher and Maddocks \cite{maddocks} who originally derived it to
		study DNA melting. This birod model is a double-stranded rod theory in which in addition to the standard variables of a Cosserat
		rod theory (i.e., center line of the rod cross-section $\mathbf{r}(s,t)$ and a material frame $[\mathbf{d}_{1}(s,t)\quad
		\mathbf{d}_{2}(s,t)\quad \mathbf{d}_{3}(s,t)]$), there are two micro-structural variables -- $\mathbf{w}(s,t)$, a micro-displacement
		measuring the change in distance between the two strands, and $\mathbf{P}(s,t)$, a micro-rotation measuring the change
		in orientation of one strand relative to the other. Fortunately, the forces conjugate to these micro-structural variables 
		obey balance laws that look similar to the balance of forces and moments equations of a standard Cosserat rod. 
		They are coupled to the macroscopic balance equations for the center-line of the rod through distributed body forces and 
		moments. Moakher and Maddocks \cite{maddocks} have provided hyper-elastic constitutive laws for these micro-structural 
		variables that are based on {\it quadratic} energies. 
		
		The theory of birods has been successfully used by Lessinnes \textit{et al.} \cite{goriely1,goriely2} to study the growth 
		and evolution of two filaments elastically bound to each other. These authors have demonstrated the utility of the theory in 
		accurately modelling biological structures across multiple length scales --from tissues and arteries, to growth of roots and 
		stems in plants. Manning \textit{et al.}~\cite{maddocks_cont} have used both a discrete base-pair model 
		and a corresponding continuum rod model to study the cyclization of short DNA molecules (150 bp). The results obtained from both 
		these approaches match remarkably well. However, for shorter length scales ($\sim 16$ bp) Lankas \textit{et al.}~\cite{maddocks_rigidbp} 
		have assayed the merit of the assumptions of rigid bases versus 
		rigid base-pairs to estimate the stiffness
		parameters of a DNA oligomer and found that the simulated data is closely consistent with the assumption of rigid bases, but not rigid
		basepairs. This inevitably necessitates the inclusion of elasticity of base-pairs via the webbing in an elastic birod \cite{maddocks} 
		to accurately model the local deformations caused by proteins at small length scales. 
		\section{Strategy to compute interaction energy} \label{sec:sec2}
		In this section, we give a concise blueprint of our strategy to approach the problem of calculating the interaction energy for two proteins binding to DNA. We model the DNA as a helical birod \cite{maddocks} which has two elastic components -- outer strands and the connecting web. The two sugar-phosphate backbones of DNA correspond to the outer elastic strands which interact by means of complimentary base-pairing represented by the elastic web in our case. We give a stepwise procedure to do the calculation and in the following sections we label each step.  
		\begin{enumerate}
			\item We begin by assuming a form of displacement for each of the outer strands which are assumed to be inextensible and unshearable.\\
			\item  We then use this displacement to calculate the tangent, normal and binormal to the deformed configuration of the outer strands thereby obtaining the rotation matrix attached to the deformed configuration of the outer strands.\\
			\item Once we get the deformation and rotation of the outer strands, we use these to calculate the extension, shear and rotation of the web.\\
			\item  At this point, we are in a position to substitute these quantities into the balance laws for the birod. We, then, seek non-zero solutions to the resulting system of differential equations. This
			leads to an eigenvalue problem.\\
			\item In the next step, we apply the boundary conditions to evaluate the constants.\\
			\item We carry out this process first when there is a single protein binding onto the DNA, and second when there are two proteins binding.\\
			\item Finally, we subtract the two energies obtained in the previous step to get our energy of interaction. We find that it takes the form of a decaying exponential oscillating with the periodicity of the underlying DNA helix.
		\end{enumerate}
		
		\section{Exponential decay of interaction energy in a `ladder'} \label{sec:sec3}
\begin{figure}[]
			\centering{
				\includegraphics[scale=0.35]{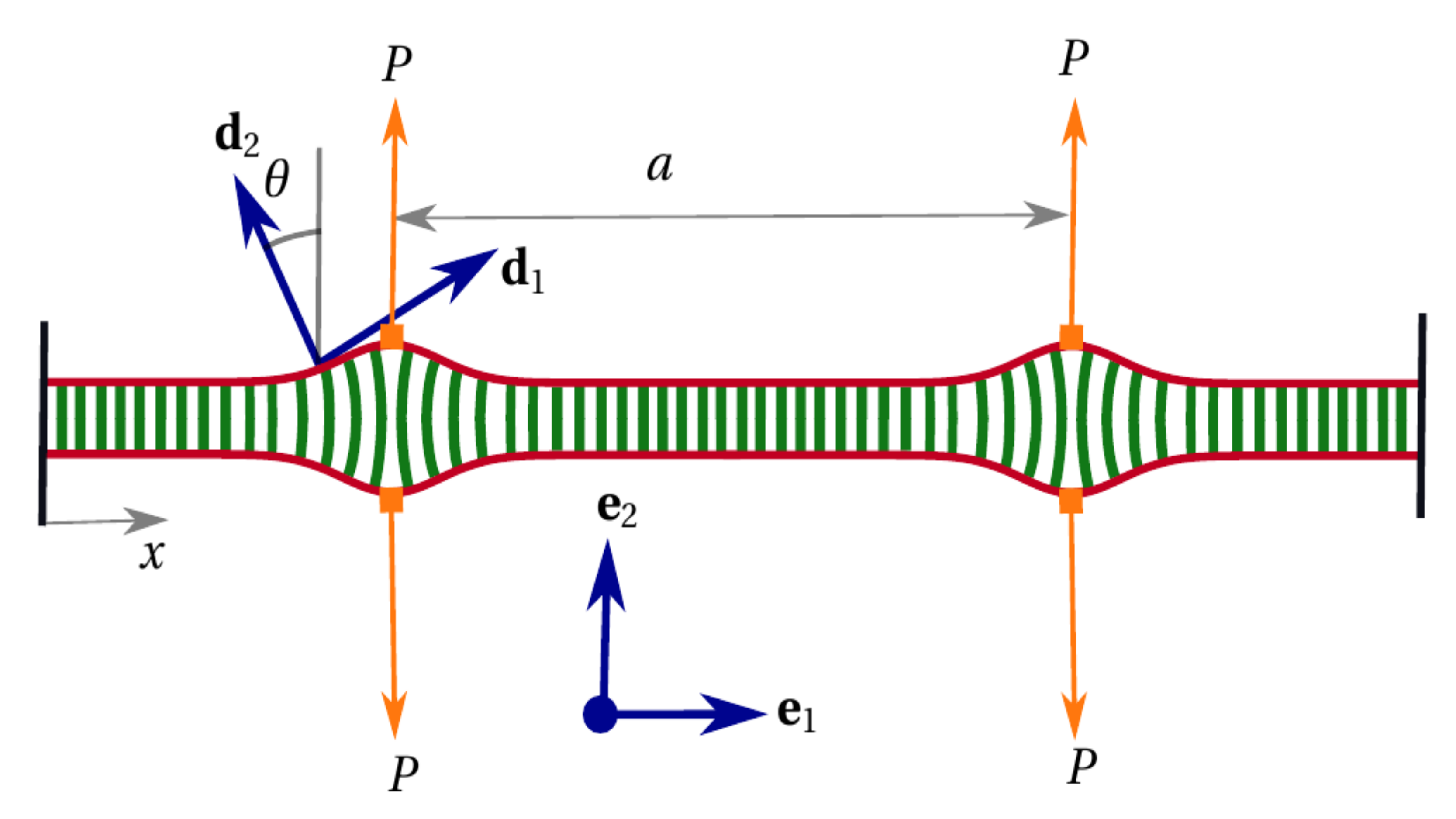}
				\caption{A straight birod, referred to as a ladder, being pulled by two force pairs separated by a distance $a$. We show that the interaction energy between the two force pairs given by eqn. \ref{Eq:int_energy} decreases exponentially with $a$. }
				\label{fig_ladder}
			}
		\end{figure}

		The calculation described above is considerably involved, so we first illustrate the main concepts in a simpler birod model which we call a `ladder' because it is not helical. We mimic the binding of a protein
		by force pairs that tend to widen the ladder as shown in fig. \ref{fig_ladder}. Our goal in this section is to demonstrate the utility of the apparatus in section \ref{sec:sec2} by computing the interaction energy for two force pairs separated by a distance $a$ as shown in fig. \ref{fig_ladder}. We work with a planar 2D birod in this section and assume small elastic deformations in the outer strands and web to keep the calculations tractable. We, ultimately, find that the interaction energy between the force pairs decays exponentially with distance $a$. 
		
		\subsection{Step 1: Kinematic description of the two strands}
		We use the arclength  parameter $x$ to describe the mechanics of the birod. In the reference configuration, both the strands $\pm$ are straight, $\br_0^\pm=x\be_1\pm\frac{d}{2}\be_2$, separated by distance $d$. Here $\be_{1}$ is a unit vector along the length of the birod, $\be_{2}$ is a unit vector perpendicular to each birod bridging the gap between them and $\be_{3}$ is normal to the plane of the birod as shown 
		in fig.~\ref{fig_ladder}. We begin by assuming a general displacement in $\be_1-\be_2$ plane. For the geometry shown in fig.~\ref{fig_ladder} we expect a mirror symmetry for deformation profiles along 
		$\be_1$ such that
		\eqsp{
			&\brp=x\be_1+\frac{d}{2}\be_2+u\be_1+w\be_2,\\
			&\brn=x\be_1-\frac{d}{2}\be_2+u\be_1-w\be_2, 
		}
		where $u=u(x)$ and $w=w(x)$ are displacements along the $\be_{1}$ and $\be_{2}$ directions, respectively.
		
		\subsection{Step 2: Rotation of the two strands}
		At each point $x$ on the $\pm$ strands we attach an orthogonal rotation frame which is simply $\bR_0^\pm=[\be_1\quad\be_2\quad\be_3]=\bI_{3\times3}$ (the identity matrix) in the reference configuration. 
		The vectors $\be_1$ and $\be_2$ map onto $\bd_{1,2}\poss$ and $\bd_{1,2}\negs$ in the deformed configuration for the positive and negative strand, respectively. The $\bd_{i}$, $i=1,2,3$ are again unit vectors. 
		\begin{equation}
		\begin{split}
		& \bd_1^\pm=\cos\theta\be_1\pm \sin\theta\be_2\approx \be_1\pm \theta\be_,\\
		&\bd_2^\pm=\mp\sin\theta\be_1+\cos\theta\be_2\approx\pm\theta\be_1+\be_2,\\
		& \bR^\pm=\begin{bmatrix}
		\cos\theta &\mp\sin\theta &0\\
		\pm\sin\theta & \cos\theta & 0\\
		0 &0 &1
		\end{bmatrix}\approx\begin{bmatrix}
		1 &\mp\theta &0\\
		\pm\theta & 1 & 0\\
		0 &0 &1
		\end{bmatrix}.
		\end{split}
		\end{equation}
		We assume small $\theta$ to keep the calculations tractable. 
		
		\subsection{Step 3: Extension and rotation of the web}
		We decompose the kinematics of the web into a macroscopic deformation and a microscopic deformation \cite{maddocks}. The former describes the rigid displacement and rotation, while the latter is related to the force and moment transferred by the web. The macro- displacement vector $\br$ is defined as $\br=\frac{\brp+\brn}{2}=x\be_1+u\be_1$ \cite{maddocks}. The macro- rotation tensor is $\bR$ defined as $\bR=(\bRp\bR^{-T})^{1/2}\bRn$ \cite{maddocks}, which in our case is 
		\eqsp{
			\bR=(\bRp\bR^{-T})^{1/2}\bRn=\mathbf{I}_{3\times3}.
		} 
		We define another tensor $\bP$ relating $\bRp$ and $\bRn$ to $\bR$. An elastic constitutive relation discussed in further sections connects the micro- rotation tensor $\bP=(\bR\poss\bR^{-T})^{1/2}$ to the moment transferred by the web.  
		\begin{equation}
		\begin{split}
		& \bP=(\bRp\bR^{-T})^{1/2}=\begin{bmatrix}
		\cos\theta &-\sin\theta &0\\
		\sin\theta & \cos\theta & 0\\
		0 &0 &1
		\end{bmatrix}\approx\begin{bmatrix}
		1 &-\theta &0\\
		\theta & 1 & 0\\
		0 &0 &1
		\end{bmatrix}.\\
		\end{split}
		\end{equation}
		We need to calculate the Gibbs rotation vector $\biet=\tan \frac{\lambda}{2}\bhk$, where $\lambda$ is obtained from $1+2\cos \lambda=\text{tr}(\mathbf{P})$ and $\bhk$ is the eigenvector of $\bP$ \textit{i.e.} $\bP\bhk=\bhk$. We need $\biet$ in the subsequent section to compute the moment transferred by the web \cite{maddocks}. By direct observations, $\lambda=\theta$ and $\bhk=\be_3$, so that $\biet=\tan \frac{\theta}{2}\be_3$. The Gibbs rotation vector in the reference configuration $\biet_0=0$.
		
		The micro- displacement of the web is defined by $\biw=\frac{\brp-\brn}{2}$, which is $\biw_0=\frac{d}{2}\be_2$ in the reference configuration and $\biw=(\frac{d}{2}+w)\be_2$ in the current configuration. We 
		need $\biw$ and $\biw_0$ to compute the force transferred by the web.  
		
		\subsection{Step 4: Governing differential equations}
		We calculate various strains and curvatures associated with the deformation and relate them to the contact force and moment, respectively, which go into the governing equations. For detailed discussion on the relations used in this section we refer the reader to Moakher and Maddocks~\cite{maddocks}. The governing equations of the birod consist of three kinetic components: the contact forces in the two strands $\bin^\pm$, the contact moments $\bim^\pm$, and the force $\bfo$ and moment $\bc$ transferred by the $-$ strand onto the $+$ strand. We compute each of these components as follows:  
		\begin{enumerate}
			\item $\bin^\pm$: We need strains in the current configuration $\biv^\pm$ and in the reference configuration $\biv^\pm_0$, in the strands to compute $\bin^\pm$. These strains are:
			\eqsp{
				&\biv^\pm_0=\pardev{\br^\pm_0}{x}=\be_1,\\
				&\biv^\pm=\pardev{\br^\pm}{x}=(1+u_x)\be_1\pm w_x\be_2.
			}  
			The contact forces $\bin^\pm=\bR^\pm \mathbf{C} \bR^{\pm T} \biv^\pm$ where $\mathbf{C}$ is a second order tensor such that $\mathbf{C}_{11}=EA$, $\mathbf{C}_{22}=GA$ and $\mathbf{C}_{12}=\mathbf{C}_{21}=0$. Here $E$ is the stretch modulus, $G$ shear modulus and $A$ is the cross-sectional area of the strands. Upon performing the calculation and taking account of the fact that $u,w$ and $\theta$ are small and upon ignoring higher order terms we get,
			\eqsp{
				\bin^{\pm}=EAu_x\be_1\pm GA(w_x-\theta)\be_2.
				\label{force_eqn}
			}
			\item $\bim^\pm$: For calculating the contact moments $\bim^\pm$ in the respective strands we need the curvature vector $\pmb{\kappa}^\pm$ for the two strands, which can, in turn, be obtained by computing the axial vector of the skew-symmetric matrices $\bU^\pm=\pardev{\bR^\pm}{x}\bR^{\pm T}$. 
			\begin{equation}
			\begin{split}
			&\bU^\pm=\pardev{\bR^\pm}{x}\bR^{\pm T}=\begin{bmatrix}
			0 & \mp\theta_x& 0\\
			\pm\theta_x & 0 & 0\\
			0 & 0 & 0
			\end{bmatrix},\\
			&\pmb{\kappa}^\pm=\pm\theta_x\be_3.
			\end{split}
			\end{equation}
			The contact moment $\bim^\pm$ is related to the curvature via a bending rigidity $EI$ such that
			\eqsp{\bim^\pm=\pm EI\theta_x\be_3.} Here, $I$ is the moment of inertia of the cross-section of the outer strands.
			\item $\bfo$ and $\bc$: The force transferred by the web $\bfo$ is proportional to the change in the dimensions of the web quantified by $\biw$ and $\biw_0$ in the previous sections such that,  
			\begin{equation}
			\begin{split}
			2\bfo=&\bR\mathbf{H}\bR^T[\biw-\bR\check{\biw_0}]\approx Lw\be_2, \\
			\end{split}
			\end{equation}
			where $\mathbf{H}$ is a diagonal second order elasticity tensor such that $\mathbf{H}_{22}=L$. Similarly, the moment transferred by the web $\bc$ is elastically related to $\boeta$ and $\boeta_0$ calculated in the previous sections:
			\begin{equation}
			\begin{split}
			2\bc&=\frac{1}{\alpha}\bR\mathbf{G}\bR^T(\biet-\bR\check{\bet})-\biet\times(\biw\times\bfo)\approx K\theta\be_3,
			\end{split}
			\end{equation}
			where $\alpha=\frac{2}{1+||\biet||^2}$, $\mathbf{G}$ is a second order diagonal elasticity tensor and $K=\frac{G_{33}}{2}$. 
		\end{enumerate}
		The governing equations from box 4 in \cite{maddocks} are given by,
		\begin{subequations}
			\begin{equation}
			\begin{split}
			& \bin_x=0,\\
			& \bim_x+\br_x\times\bin=0.
			\end{split}
			\end{equation}
			\begin{equation}
			\begin{split}
			& \binc_x-2\bfo=0,\\
			& \bimc_x+\br_x\times\binc-\bc=0.
			\end{split}
			\end{equation}
		\end{subequations}
		In the above equations, $\bin=\binp+\binn=2EAu_x\be_1$, $\binc=\binp-\binn=2GA(w_x-\theta)\be_2$, $\bim=\bimp+\bimn+\biw\times\binc=0$ and $\bimc=\bimp-\bimn+\biw\times\bin=2EI\theta_x\be_3+(d/2+w)\be_2\times 2EAu_x\be_1\approx2[EI\theta_x-\frac{d}{2}EAu_x]\be_3$. Upon substituting these values into the governing equations we get,
		\begin{equation}
		\begin{split}
		& EAu_{xx}=0,\\
		& 2GA(w_{xx}-\theta_x)-Lw=0,\\
		& 2(EI\theta_{xx}-d/2EAu_{xx})+(1+u_x)\be_1\times2GA(w_x-\theta)\be_2-K\theta\be_3=0.
		\end{split}
		\end{equation}
		We use $\theta_x=w_{xx}-\frac{L}{2GA}w$ and $u_{xx}=0$ and get,
		\begin{equation}
		EIw_{xxxx}-(\frac{EIL}{2GA}+\frac{K}{2})w_{xx}+(\frac{L}{2}+\frac{KL}{4GA})w=0.
		\end{equation}
		If we further assume that the outer strands are unshearable ($GA\to \infty$ and $\theta=w_x$), the above equation reduces to a simpler equation.
		\eqsp{EIw_{xxxx}-\frac{K}{2}w_{xx}+\frac{L}{2} w=0.
			\label{eq:gov_eqn_ladder}
		}
		
		\subsection{Step 5,6 and 7: Interaction Energy}
		We substitute $w=e^{ms}$, and get eigenvalues $m=\pm \lambda,\pm \mu$. For illustration purposes, we assume $\lambda$ and $\mu$ are real numbers (i.e., $K^2-32L>0$) and the ladder extends from $-\infty$ in the negative $\be_{1}$ direction to $+\infty$ in the positive $\be_{1}$ direction with $w=w_x=0$ at $x=\pm\infty$. Hence, for a force pair at $x=0$
		\eqsp{
			& w(x)=Ae^{\lambda x}+Be^{\mu x}\quad\text{when}\quad x<0,\\
			& w(x)=Ae^{-\lambda x}+Be^{-\mu x}\quad\text{when}\quad x>0,
		}
		for some constants $A$ and $B$ which could be determined using boundary conditions in step 5. For two force pairs separated by a distance $a$, the displacement profile $w_2(x)=w(x)+w(x-a)$. The elastic energy in the deformed configuration is computed in step 6 and is given by,
		\eqsp{
			E[w]=EI w_{xx}^2+\frac{1}{2}Kw_x^2+\frac{1}{2}Lw^2.
		}
		Finally, we compute the interaction energy defined by $\Delta G=E[w_2]-2E[w]$ in step 7 and find that it decreases exponentially with the distance $a$.
		\eqsp{
			\Delta G=&\frac{L}{2} \Big(\frac{e^{-\lambda a} \left(A^2 \lambda ^2 \mu -A^2 \mu ^3+A^2 \lambda ^3 \mu  a-A^2 \lambda  \mu ^3 a-4 A B \lambda  \mu ^2\right)}{\lambda  \mu  (\lambda^2 -\mu^2 )}
			+\\
			&\quad\quad\frac{e^{-\mu a} \left(4 A B \lambda ^2 \mu +B^2 \lambda ^3-B^2 \lambda  \mu ^2+B^2 \lambda ^3 \mu  a-B^2 \lambda  \mu ^3 a\right)}{\lambda  \mu  (\lambda^2 -\mu^2 )}\Big)+\\
			&\frac{K}{2}\Big(\frac{e^{-\lambda a} \left(A^2 \lambda ^3-A^2 \lambda  \mu ^2-A^2 \lambda ^4 a +A^2 \lambda ^2 \mu ^2 a+4 A B \lambda ^2 \mu \right)}{(\lambda^2 -\mu^2 )}+\\
			&\quad\quad\frac{e^{-\mu a } \left(-4 A B \lambda  \mu ^2+B^2 \lambda ^2 \mu -B^2 \mu ^3-B^2 \lambda ^2 \mu ^2 a+B^2 \mu ^4 a\right)}{(\lambda^2 -\mu^2 )}\Big)+\\
			& EI\Big(\frac{e^{-\lambda a} \left(A^2 \lambda ^5-A^2 \lambda ^3 \mu ^2+A^2 \lambda ^6 a-A^2 \lambda ^4 \mu ^2 a-4 A B \lambda ^2 \mu ^3\right)}{(\lambda^2 -\mu^2 )}+\\
			&\quad\quad\frac{e^{-\mu a} \left(4 A B \lambda ^3 \mu ^2+B^2 \lambda ^2 \mu ^3-B^2 \mu ^5+B^2 \lambda ^2 \mu ^4 a-B^2 \mu ^6 a\right)}{(\lambda^2 -\mu^2 )}
			\Big).
		}
		In the next section we will follow these steps for a helical birod model of DNA.
		
		\section{Interaction energy for two DNA binding proteins}
		\subsection{Assumptions}
		We first outline the various assumptions and set up the underlying framework for our helical birod model of DNA. We assume elastic deformations throughout. When a protein binds to DNA it causes local bending and twisting. We assume that the resulting twist and curvatures are small. These curvatures could possibly add up to produce large displacements and rotations. The two phosphate backbones of DNA constitute the helical outer strands which are out of phase by a phase angle $\alpha=2.1$ radians. We assume these backbones to be inextensible. These outer strands consist of sugar phosphate single bonds. Thus, we assume that they can not support twisting moments. The inextensibility of the outer strands is a strong geometrical constraint which induces a change in the radius and phase angle between the two helices when a protein 
		causes local deformations. We assume that these changes are small and of the same order as the curvatures.    
		
		\subsection{Step 1: Deformation of the outer strands}
\begin{figure}[]
			\centering{
				\includegraphics[scale=0.49]{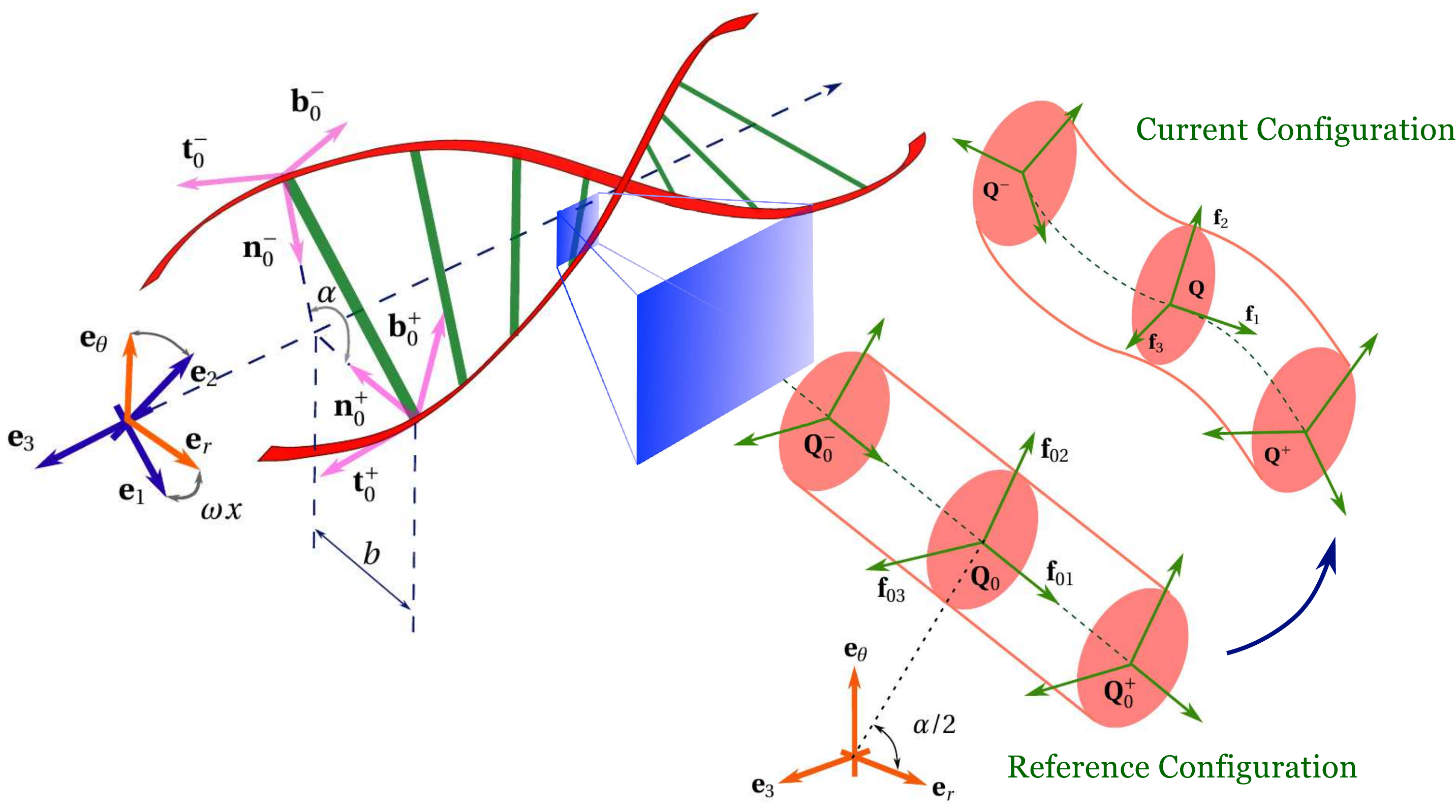}
				\caption{A DNA molecule as a double helical elastic birod is shown on the left. The phosphate backbones are represented by outer strands while the complimentary base-pairing is represented by the elastic web. The phase angle between the two helices is $\alpha=2.1$ radians. Here $\bRp=[\npz\quad\bpz\quad\tpz]$ and $\bRn=[\nnz\quad\bnz\quad\tnz]$ are the Frenet-Serret frames attached to the $+$ and $-$ strands, respectively. Base-pairs in reference and current configuration are shown to the right. $\bQ\poss_0=\bQ\negs_0=\bQ_0$ in the referenece configuration. In the current configuration, the rigid rotation of the base-pair is quantified by $\bQ=\bZ(\bI+\bphi)\bQ_0$ (eqn. \ref{eq:Q}) and the elastic moment $\bc$ is related to the Gibbs rotation vector of  $\bP=(\bQ\poss\bQ^{-T})^{\frac{1}{2}}$ (eqn. \ref{eq:P}).}
				\label{fig_birod}
			}
		\end{figure}

		DNA consists of two helical strands with radius $b=1$ nm and pitch $p=3.4$ nm, out of phase by $\alpha=2.1$ radians, wrapped around a common axis as shown in fig. \ref{fig_birod}. We follow the notation used by Moakher and Maddocks \cite{maddocks} and refer to the two stands as $\pm$. The undeformed state of the outer strands denoted by $\br_0^\pm(x)$ is a helix with a constant radius and pitch. We choose to parametrize both the curves by arclength parameter $x$. Here, $\omega=\frac{2\pi}{p}$ and $k$ is the characteristic angle of the helix such that $\tan k =\frac{2\pi b}{p}=\omega b$. 
		\[
		\brp_0=b\Big(\cos \omega x\be_1+\sin \omega x\be_2\Big)+x\be_3,
		\]
		\[
		\brn_0=b\Big(\cos (\omega x+\alpha)\be_1+\sin (\omega x+\alpha)\be_2\Big)+x\be_3.
		\]
		Let us now focus on the two strands separately. The calculations for the $+$ strand are given in this section while the results for the $-$ strand are given in the appendix. We posit a form of displacement wherein the radius of the helix changes and its axis is allowed to take arbitrary shapes within the ambit of the assumptions specified in section \ref{sec:sec2}. Here $[\be_1,\be_2,\be_3]$ denotes the standard spatial reference frame and $\be_3$ is along the common axis of the two helices $\pm$ in the \textit{reference} configuration. This common axis in the \textit{deformed} configuration is defined by the set of orthogonal directors $[\bd_1(x),\bd_2(x),\bd_3(x)]$. The displacement fields which define the undeformed and deformed configuration are, 
		\begin{equation}
		\begin{split}
		&\brp_0=b\Big(\cos \omega x\be_1+\sin \omega x\be_2\Big)+x\be_3,\\
		&\brp(x)=(b+r)\Big(\cos(\omega x+\beta^+)\bd_1+\sin(\omega x+\beta^+)\bd_2\Big)+\int_{0}^{x}dx(1+b\xi)\bd_3.\\
		\end{split}
		\label{eq_r_def}
		\end{equation}
		where
		\[
		r=r(x),\quad \betap=\betap(x),\quad \xi=\xi(x).\]
		Here $r$ is the change in the radius of the helix, $\betap$ is the change in the phase of the $+$ strand, and $\xi$ can be considered as a stretching of the axis of the helix. 
		
		Let $\bZ$ be a second order orthogonal tensor which relates the directors of the deformed centerline $\bd_i$ to those of the undeformed one $\be_i$, $i=1,2,3$. As stated in section 2, the curvatures $(k_1,k_2,k_3)$ associated with the deformation of the centerline are assumed to be small, nonetheless these could aggregate to potentially produce large rotations. The orthogonal tensor $\bZ$ operates as follows.
		\eqsp{\bd_i=\mathbf{Z}\be_i,\quad\quad \bZ=\sum_{i=1}^{3}\bd_i\otimes\be_i,\quad \quad i=1,2,3\\
			\label{def_Z}
		}
		and
		\begin{equation}
		\begin{split}
		& \bd_{ix}=\boldsymbol{\upkappa}\times\bd_i, \quad\quad\text{where}\quad \pmb{\upkappa}=k_1\bd_1+k_2\bd_2+k_3\bd_3.\\
		&\bd_{1x}=k_3\bd_2-k_2\bd_3,\quad\quad\bd_{2x}=k_1\bd_3-k_3\bd_2,\quad\quad\bd_{3x}=k_2\bd_1-k_1\bd_2.\\
		\end{split}
		\end{equation}
		In the above equations, we assume that
		\[ r^+(x),k_1(x),k_2(x),k_3(x),\zeta(x), \beta^+(x)\sim O(\varepsilon).\]
		Thus, in the treatment henceforth, any product terms such as $r^2$ or $\xi k_3$ are $O(\varepsilon^2)$ and are neglected. 
		
		\subsection{Step 2: Rotation of strands}
		We proceed in a standard way by attaching a Frenet-Serret director frame consisting of normal, binormal and tangent to each cross-section of the strand as shown in fig.~\ref{fig_birod}. We denote it 
		by $\bRp_0(x)$ in the reference configuration. 
		\begin{equation}
		\begin{split}
		&\bRp_0=[\np_0\quad\bp_0\quad\tp],\\
		&\npz=-\cosr \be_1-\sinr \be_2,\\
		&\bpz=-\cos k (-\sinr \be_1+\cosr\be_2)+\sin k\be_3,\\
		&\tpz=\sin k(-\sinr \be_1+\cosr\be_2)+\cos k \be_3,
		\end{split}
		\end{equation} 
		For the sake of brevity, we use
		\[
		\bfone=\bfv\poss_1,\quad \bftwo=\bfv\poss_2,\quad \bd_3=\bfv\poss_3.
		\]
		As the strand deforms, the frame $\bRp_0$ evolves into $\bRp(x)$ which consists of normal, binormal and tangent to the deformed configuration of the strand. Our next step is to calculate the tangent vector to the deformed configuration. We differentiate eqn. (\ref{eq_r_def}) to obtain,
		\begin{equation}
		\begin{split}
		\brp_x&=(r_x-b\omega \beta^+)\bfone+(b\omega +\omega r +b\beta^+_x+bk_3)\bftwo+\\
		&(1+b\xi-bk_2\cosr+bk_1\sinr)\bd_3.
		\end{split}
		\label{eq_rx}
		\end{equation}
		We assume the strand to be inextensible and unshearable. This means, 
		\[
		|\brp_x|^2=1+\omega^2b^2+2b(\omega^2 r +b\omega \beta^+_x+b\omega k_3+\xi-k_2\cosr+k_1\sinr )+O(\varepsilon^2)=|\brp_{0x}|^2=1+b^2\omega^2,
		\]
		which leads us to the inextensibility condition: 
		\begin{equation}
		\xi-k_2\cosr+k_1\sinr=-\omega^2 r-b\omega(k_3+\beta^+_x).
		\label{eq:inext}
		\end{equation}
		We will subsequently use this equation to impose boundary conditions. We substitute eqn. (\ref{eq:inext}) into into eqn. (\ref{eq_rx}) to get, 
		\begin{equation}
		\begin{split}
		\brp_x&=(r_x-b\omega \beta^+)\bfv\poss_1+(b\omega +\omega r +b\beta^+_x+bk_3)\bfv\poss_2+(1-b\omega^2 r-b^2\omega(k_3+\beta^+_x))\bfv\poss_3.
		\end{split}
		\end{equation}
		Now, we need to find the director frame for the strand in the deformed configuration. We start by calculating the tangent vector,
		\begin{equation}
		\begin{split}
		\tp=&\frac{\brp_x}{|\brp_{0x}|}\\
		=&(r_x\cos k-\beta^+\sin k )\bfv_1\poss+(\sin k+ \omega r\cos k+b(\beta^+_x+k_3)\cos k)\bfv_2\poss\\
		&+(\cos k- \omega r\sin k-b(\beta^+_x+k_3)\sin k )\bfv_3\poss\\
		=&\bZ(\tpz-(r_x\cos k-\beta^+ \sin k)\npz-(\omega r+b(\beta^+_{x}+k_3))\bpz).
		\end{split}
		\end{equation}
		We differentiate the tangent vector to calculate the normal in the deformed configuration
		\begin{equation}
		\begin{split}
		\tp_x=&(-\omega \sin k +(r_{xx}+\xi)\cos k-(\beta_{x}+k_3)\sin k)\bfv\poss_1\\
		&+(2\omega \cos k r_x+-\omega \beta^+ \sin k+b\cos k(\beta^+_{xx}+k_{3x})-f\cos k)\bfv\poss_2\\
		&+(f-\omega r_x-b(\beta^+_{xx}+k_{3x}))\sin k\bfv\poss_3+\oepss.
		\end{split}
		\end{equation}
		We can use the above expression to calculate the curvature $\Omega\poss$ for the strand. We find that this is equal to the sum of the original curvature ($\omega \sin k$) and the one induced by the process of deformation $\kappa\poss$.  Hence, 
		\begin{equation}
		\begin{split}
		\Omega \poss=&(\tp_x.\tp_x)^{1/2}=\omega\sin k-(r_{xx}+\xi)\cos k+(\beta^+_{x}+k_3)\sin k,\\
		\kappa\poss=&\Omega\poss-\omega \sin k=-(r_{xx}+\xi)\cos k+(\beta^+_{x}+k_3)\sin k.
		\end{split}
		\end{equation}
		The bending moment $\bimp$ in the strand is proportional to $\kappa\poss$. 
		\begin{equation}
		\begin{split}
		\bimp=EI \kappa\poss~\bp=EI \kappa\poss(-\cos k\bfv\poss_2 + \sin k \bfv\poss_3).\\
		\end{split}
		\label{eq:kappa}
		\end{equation}
		Also, the normal is 
		\begin{equation}
		\begin{split}
		\np=\frac{1}{\Omega\poss}\tp=&-\bfv\poss_1+\frac{1}{\sin k}(r_x \sin k-\beta^+ \sin k)\bfv\poss_2\\
		&+\frac{f-\omega r_x-b(\beta^+_{xx}+k_{3x})}{\omega \sin k}(-\cos k\bfv_2+\sin k \bfv\poss_3),\\
		&=\bZ\big(\npz+(r_x\cos k-\beta^+ \sin k)\tpz+(-\frac{(r_x\cos k-\beta^+ \sin k)\cos k}{\sin k}+\frac{g}{\omega \sin k})\bpz\big).\\
		\end{split}
		\end{equation}
		where
		\[
		g(x)=f(x)-\omega r_x-b(\beta^+_{xx}+k_{3x}),\quad \quad f(x)=k_1\cosr+k_2\sinr.
		\]
		Using the above deformed orthogonal frame attached to each cross section 
		\begin{equation}
		\bRp=[\np\quad\bp\quad\tp]=\mathbf{Z}\bRp_0(\mathbf{1}+\thetap),
		\end{equation} 
		where $\thetap$ is a skew symmetric tensor and $\bZ=\sum_{i=1}^{3}\bd_i\times\be_i$ as defined in eqn. \ref{def_Z},
		\begin{equation}
		\begin{split}
		&\thetap=\begin{bmatrix}
		0 &-\theta^+_3&\theta^+_2\\\theta^+_3&0&-\theta^+_1\\-\theta^+_2&\theta^+_1&0
		\end{bmatrix},\\
		\text{in which}\quad\quad& \theta^+_1=(r\omega+b(\beta^+_x+k_3)),\quad \theta^+_2=-r_x\cos k+\beta^+ \sin k ,\\
		& \theta^+_3=\frac{g}{\omega \sin k}-\frac{(r_x\cos k-\beta^+ \sin k)\cos k}{\sin k}.
		\end{split}
		\label{eq:Q0}
		\end{equation}
		We can derive all the above quantities $\brn, \bRn$ and $\kappa\negs$ etc., for the $-$ strand too. We give the relevant expressions for these quantities in the appendix.  
		
		\subsection{Step 3: Mechanics of base-pairing}
		The sugar-phosphate backbones of the DNA molecule are tied together by means of complimentary base-pairing. We model the base-pairing by elastic rods capable of extension, shear, bending and twisting. We attach the orthogonal frame $\bQ_0=[\bfv_{01}\quad\bfv_{02}\quad\bfv_{03}]$ to the strands such that $\bfv_{01}$ is a unit vector pointing from the $-$ strand to the $+$ strand in the reference configuration as 
		shown in fig.~\ref{fig_birod}. Thus, 
		\begin{equation}
		\begin{split}
		& \mathbf{Q_0}=[\mathbf{f}_{01}\quad\mathbf{f}_{02}\quad\mathbf{f}_{03}],\\
		\quad\bfv_{01}=\sinrabt\be_1-\cosrabt\be_2,&\quad\bfv_{02}=\cosrabt\be_1+\sinrabt\be_2,\quad\bfv_{03}=\be_3.
		\end{split}
		\end{equation} 
		
		We denote the two ends of the rod in the web as $\pm$ such that the $+$ end lies on the $+$ strand and the $-$ end lies on the $-$ strand. The deformation of the web is completely determined by the displacement $(\brp(x),\brn(x))$ and rotation $(\bRp(x),\bRn(x))$ of its ends.  As the outer strands undergo the deformation prescribed by eqn. (\ref{eq_r_def}), the strands themselves undergo various kinds of deformation. We describe the rotation of the web via a rigid rotation and a micro-rotation \cite{maddocks}. The micro-rotation encapsulates the information about the difference in rotation of the two ends of the web. We 
		calculate the mechanical quantities associated with the extension and bending of the web in two separate sections below.
		
		\subsubsection{Bending and twisting of the web}
		Our objective in this section is to calculate the micro-rotation tensor $\bP$. We attach a copy of $\mathbf{Q}_0$ say $\mathbf{Q_0^\pm}$ on the + and - end of every spoke in the reference configuration. $\bQ_0^\pm$ change to $\bQ^\pm$ in the current configuration. The 'difference' between $\mathbf{Q_0^+}$ and $\mathbf{Q_0^-}$ gives the bending and torsion of the web while the 'average' of $\mathbf{Q_0^+}$ and $\mathbf{Q_0^-}$ gives the rigid rotation of the web. We relate $\mathbf{Q^\pm}$ to the rotations of $\pm$ strands $\bR(x)^\pm$. The angles between the columns of $\bQ^+_0$ and $\bRp_0$ should remain same during the deformation which translates into the following condition.
		\begin{equation}
		\begin{split}
		&\bR_0^{+T}\bQ_0=\bR^{+T}\bQ^+,\\
		&\bQ^+=\bRp\bR_0^{+T}\bQ_0=\bZ\bRp_0(\mathbf{1}+\thetap)\bR_0^{+T}\bQ_0,	\\
		&\bQ^-=\bZ\bRn_0(\mathbf{1}+\thetan)\bR_0^{-T}\bQ_0.
		\end{split}
		\end{equation}
		We are now in a position to calculate the micro-rotation $\bP$ responsible for generating elastic moment in the web. Let the micro-rotation tensor in the reference configuration be $\bP_0$ which changes to 
		$\bP$ during deformation. We use an expression for $\bP/\bP_0$ given in Moakher and Maddocks \cite{maddocks}.  
		\begin{equation}
		\begin{split}
		\bP_0^2=&\bQ_0\poss \bQ_0^{-T}=\mathbf{I},\\
		\bP^2=\bQ^+\bQ^{-T}=&\bZ\bRp_0(\mathbf{1}+\thetap)\bR_0^{+T}\bQ_0\bQ_0^T\bRn(\mathbf{1}-\thetan)\bR_0^{-T}\bZ^T,\\
		=&\bZ(\bI+\bRp\thetap\bR^{+T}-\bRn\thetan\bR^{-T})\bZ^T.\\
		\end{split}
		\end{equation}
		This gives
		\begin{equation}
		\bP_0=\mathbf{I},\quad\quad\bP\approx\bZ(\bI+\frac{\bRp\thetap\bR^{+T}-\bRn\thetan\bR^{-T}}{2})\bZ^T=\bZ(\bI+\bphic)\bZ^T.
		\label{eq:P}
		\end{equation}
		Note that $\bphic$ is a skew symmetric tensor. The next step is to calculate the Gibbs rotation vector of $\bP$ \cite{maddocks}. The Gibbs rotation vector $\bar{t}$ of a rotation matrix $\mathbf{T}$ is defined as $\bar{t}=\tan \frac{\theta}{2} \mathbf{k}$ such that $\text{tr}\bP=1+2\cos \theta$ and $\mathbf{k}$ is a unit vector such that $\mathbf{T}\mathbf{k}=\mathbf{k}$. Consider $\bar{\bP}=\bI+\bphic$ where $\bphic\sim \oeps$. The axis of the infinitesimal rotation $\bar{\bP}$ is the axial vector of $\bphic$. Hence, 
		\begin{equation}
		\bar{\bP}\phic=(\bI+\bphic)\phic=\phic,\quad\quad\text{which gives}\quad\mathbf{k}=\frac{\phic}{|\phic|}.
		\end{equation}  
		We can not calculate the magnitude of the rotation by taking $\text{tr}\bar{\bP}$, since it gives $1+2\cos \theta=3$ which implies $\theta=0$. We consider the following limit.
		\begin{equation}
		1+\bphic=\lim_{\phi^c_1\to 0}\lim_{\phi^c_2\to 0}\lim_{\phi^c_3\to 0}\bR_1(\phi^c_1)\bR_2(\phi^c_2)\bR_3(\phi^c_3).
		\end{equation}
		Now we take the trace of the RHS and get $\theta=|\phic|$. Hence, the Gibbs rotation vector of $\bar{\bP}$, $\etabar$ is given as
		\begin{equation}
		2\etabar=2\tan \frac{\theta}{2} \mathbf{k}\approx|\phic|\frac{\phic}{|\phic|}=\phic.
		\end{equation}
		The Gibbs rotation vector of $\bP$ is simply $\boeta=\bZ\etabar$. Note that in the undeformed state $\boeta_0=\etabar_0=0$.
		We now proceed to calculate the rigid rotation of the spoke $\bQ$. 
		\begin{equation}
		\bQ=\bP\bQ^-=\bZ(\bI+\frac{\bRp\thetap\bR^{+T}+\bRn\thetan\bR^{-T}}{2})\bQ_0=\bZ(\bI+\bphi)\bQ_0=\bZ(\bI+\bphi)\bQ_0.
		\label{eq:Q}
		\end{equation}
		Here $\boeta\sim O(\varepsilon)$. Now, the micro-moment $\bc$ is related linearly to the $\boeta$ via an elastic tensor $\mathbf{H}$.
		\begin{equation}
		\begin{split}
		\bc=&\bQ\mathbf{\bar{H}}[\bQ^T\boeta-\bQ_0^T\boeta_0]+\oepss\approx\bZ\bQ_0\mathbf{\bar{H}}\bQ_0^T\etabar.
		\end{split}
		\label{eq:micromoment}
		\end{equation}
		For further reference, let 
		\eqsp{
			\hat{\pmb{\zeta}}=\bQ_0^T\bar{\eta}.
			\label{eq:microbending}
		}
		\subsubsection{Extension of the web}
		The distance between the two strands is $\biw=\frac{\brp-\brn}{2}$ and in the undeformed configuration $\bw_0=\frac{\brp_0-\brn_0}{2}$. By direct calculation we observe
		\begin{equation}
		\begin{split}
		\bw_0=&b\sinabt\Big(\sinrabt\be_1-\cosrabt\be_2\Big),\\
		\biw=&(b\sinabt+w_1)\bfonet+w_2\bftwot,\\
		\end{split}
		\label{eq:microstrain}
		\end{equation} 
		where
		\[
		w_1=\frac{r+r\negs}{2}\sinabt-b~\frac{\beta\poss-\beta\negs}{2}\cosabt,\quad\text{and}\quad w_2=\frac{r-r\negs}{2}\cosabt+b~\frac{\beta\poss+\beta\negs}{2}\sinabt.
		\]
		The force exerted by the $+$ strand on the $-$ strand $\bfo$ is given by,
		\begin{equation}
		\begin{split}
		\bfo=&\bQ\mathbf{\bar{L}}[\bQ^T\biw-\bQ_0^T\biw_0],\\
		\end{split}
		\label{eq:microforce}
		\end{equation}
		where $\mathbf{\bar{L}}$ is a tensor of mechanical properties of the web. This force $\bfo$ causes the web to extend and shear.
		For further reference let,
		\eqsp{
			\hat{\bw}=\bQ^T\biw-\bQ_0^T\biw_0
		} 
		\subsubsection{Stacking energy} 
		DNA consists of consecutive base-pairs stacked on top of each other in a regular fashion. The resistance to external forces and moments not only comes from the elastic deformation of the strands and the 
		webbing but also from the change in alignment of the base-pairs. We call the energy associated with this change in bases' position and spatial orientation `stacking energy'.  
		Stacking energy plays a critical role in various phenomena such as melting of DNA \cite{stacking_energy,stacking_energy2}. We prescribe a form of free energy which is quadratic in the twist $k_3$ and 
		stretch $\xi$.
		\begin{equation}
		F_{int}=K_c k_3^2+K_e\xi^2.
		\end{equation} 
		There are other sophisticated expressions for the stacking energy~\cite{stacking_energy2}, but we use the quadratic form for two reasons: one, the non-quadratic terms in the energy 
		of \cite{stacking_energy2} account for effects such as base-pair severing which are crucial to DNA melting which does not occur in our problem, two, a quadratic energy keeps our problem linear.  
		This interaction energy results in a distributed body force $\boldsymbol{l}$ and distributed body moment $\boldsymbol{h}$ on the strands.
		\begin{equation}
		\boldsymbol{h}=K_c k_{3x}\bd_3,\quad\quad\boldsymbol{l}=K_e \xi_x \bd_3. \label{eq:int}
		\end{equation}
		
		\subsection{Step 4: Governing equations} \label{sec:step4}
		We are now in a position to solve the governing equations for the mechanics of our helical birod. These equations consist of balance of linear momentum and angular momentum for both the strands. In the 
		balance equations eqn. (\ref{eq:f_bal}) and eqn. (\ref{eq:m_bal}):
		\begin{itemize}
			\item $\bim^\pm=EI\kappa^\pm$ (eqn. \ref{eq:kappa}) denotes the elastic moment in the $\pm$ strand. $\bin^\pm$ are the contact forces for which there is no constitutive relation since the outer strands are assumed to be inextensible and unshearable.
			\item $\bfo$ and $\bc$ are the distributed force and moment, respectively, exerted by the $+$ strand on the $-$ strand.  
			\item $\boldsymbol{l}$ and $\boldsymbol{h}$ are the distributed force and moment exerted by base-pairs on the $+$ and $-$ strand.
		\end{itemize} 
		The balance equations are:
		\begin{subequations}
			\begin{equation}
			\binp_x-\bfo+\boldsymbol{l}=0,
			\end{equation}
			\begin{equation}
			\binn_x+\bfo+\boldsymbol{l}=0,
			\end{equation}
			\label{eq:f_bal}
		\end{subequations}
		\begin{subequations}
			\begin{equation}
			\bimp_x+\brp_x\times\binp+\frac{1}{2}(\brp-\brn)\times\boldsymbol{f}-\boldsymbol{c}+\boldsymbol{h}=0,
			\end{equation}
			\begin{equation}
			\bimn_x+\brn_x\times\binn+\frac{1}{2}(\brp-\brn)\times\boldsymbol{f}+\boldsymbol{c}+\boldsymbol{h}=0,
			\end{equation}
			\label{eq:m_bal}
		\end{subequations}
		Let $[\bfv_1\quad\bfv_2\quad\bfv_3]=\bZ\bQ_0$. This gives
		\eqsp{
			&\bfv_1=\bfonet,\quad\bfv_2=\bftwot,\quad\bfv_3=\bd_3.
			\label{Eq:Q}
		}
		We decompose the forces, $\binp=(\bin+\binc)\sim \oeps$ and $\binn=(\bin-\binc)\sim \oeps$. $\bin=n_1\bfv_1+n_2\bfv_2+n_3\bfv_3$ and $\binc=n_1^c\bfv_1+n_2^c\bfv_2+n_3^c\bfv_3$.  
		Now, $\bin_x=(n_{1x}-\omega n_2)\bfv_1+(n_{2x}+\omega n_1)\bfv_2+n_{3x}\bfv_{3}+\oepss$. Similarly for $\binc_x$. We use $\bc=c_1\bfv_1+c_2\bfv_2+c_3\bfv_3$ and $\boldsymbol{f}=f_1\bfv_1+f_2\bfv_2+f_3\bfv_3$ from eqn. (\ref{eq:microforce}) and eqn. (\ref{eq:micromoment}). Then, the balance equations become:  
		\begin{equation}
		\begin{split}
		&n_{1x}-\omega n_2=0,\\
		&n_{2x}+\omega n_1=0,\\
		&n_{3x}+K_e\xi_x=0,\\
		&n^c_{1x}-\omega n^c_2-f_1=0,\\
		&n^c_{2x}+\omega n^c_1-f_2=0,\\
		&n^c_{3x}-f_3=0,\\
		&EI\cos{k}[(\kappa\poss_x+\kappa\negs_x)\cosabt+(\kappa\poss-\kappa\negs)\omega\sinabt]-2n_2+2a\omega n^c_3\sinabt=0,\\
		&EI\cos{k}[(\kappa\negs_x-\kappa\poss_x)\sinabt+(\kappa\negs+\kappa\poss)\omega\cosabt]+2n_1+2a\omega n_3\cosabt-2af_3\sinabt=0,\\
		&EI\sin k(\kappa\poss_x+\kappa\negs_x)+2af_2\sinabt-2sn_2\cosabt-2a\omega n^c_1\sinabt+2K_c k_{3x}=0,\\
		&EI\cos{k}[(\kappa\poss_x-\kappa\negs_x)\cosabt+(\kappa\poss+\kappa\negs)\omega\sinabt]+2a\omega n_3\sinabt-2n^c_2-2c_1=0,\\
		&EI\cos{k}[-(\kappa\negs_x+\kappa\poss_x)\sinabt+(\kappa\poss-\kappa\negs)\omega\cosabt]+2\omega n^c_3\cosabt+2n^c_1-2c_2=0,\\
		&EI\sin k(\kappa\poss_x-\kappa\negs_x)-2a\omega n^c_2\cosabt-2a\omega n_1\sinabt-2c_3=0,
		\end{split}
		\label{eq:12diffeqns}
		\end{equation}
		We have 12 differential equations in the 12 unknowns $(r,f,\xi,k_3,\beta\poss,\beta\negs,n^c_1,n^c_2,n^c_3,n_1,n_2,n_3)$. We substitute the following ansatz into the equations. 
		\begin{equation}
		y=y_0e^{-\lambda x} \quad \text{where $y$ could be }r(x),f(x),\xi(x),k_{3}(x),\beta\poss(x),\beta\negs(x),n^c_1,n^c_1,n^c_3,n_1,n_2,n_3.
		\end{equation}
		This results in an eigenvalue problem. We find $23$ eigenvalues, but retain only $6$ for reasons explained in the appendix.  Let those $6$ eigenvalues be $\pm\lambda,\pm \mu, \pm \delta$ and the corresponding 
		eigenvectors $\bv_{\pm\lambda}$ and $\bv_{\pm \mu}$. Let 
		\[
		\bv(x)=[r(x)\quad f(x)\quad \xi(x)\quad k_3(x)\quad \beta\poss(x)\quad\beta\negs(x)\quad n^c_1(x)\quad n^c_2(x)\quad n^c_3(x)\quad n_3(x)\quad n_1(x)\quad n_2(x)]^T.
		\]
		Hence, 
		\begin{equation}
		\bv(x)=p_1e^{-\lambda x}\bv_{\lambda}+p_2e^{\lambda x}\bv_{-\lambda}+p_3e^{-\mu x}\bv_{\mu}+p_4e^{\mu x}\bv_{-\mu}+p_5e^{-\delta x}\bv_{\delta}+p_6e^{\delta x}\bv_{-\delta}.
		\label{Eq:soln}
		\end{equation}
		Here, $p_1$, $p_2$, $p_3$, $p_4$, $p_5$ and $p_6$ are the constants which are determined using boundary conditions.
		
		\subsection{Step 5: Boundary conditions} \label{sec:step5}
		We assume that the impact of a protein binding to DNA is two fold: a) the protein fixes the curvatures at the binding site as in \cite{spakowitz,xiajun_extmech,curv_sensing_prot_goriely}, and b) the protein causes a change in the radius of the DNA helix~\cite{sunscience} as shown in the inset of fig. \ref{fig:deltaG} (b). Thus, we apply boundary conditions on the curvatures $k_1$, $k_2$ and the change in radius $r$ of the DNA helix. We discuss two cases, first, when one protein binds to the DNA, and second, when two proteins bind to it. 
		\begin{enumerate}
			\item \textit{One protein}: Let us assume that the protein binds at $x=0$. The boundary conditions for this case are:
			\begin{equation}
			\begin{split}
			\text{At}\quad&x=0,\quad \quad k_1(x)=k_{10},\quad k_2(x)=k_{20},\quad r(x)=r_0.\\
			\text{As}\quad&x\to\pm\infty,\quad\quad k_1(x),k_2(x), r(x)\to0.
			\end{split}
			\end{equation}
			The second boundary condition says that the DNA is straight far away from the protein and that the perturbation in DNA radius occurs only in the vicinity of the bound protein.
			\item \textit{Two proteins}: Let us assume that the two proteins bind at $x=0$ and $x=a$, respectively. We divide our domain into three parts $-\infty<x<0$, $0<x<a$ and $a<x<\infty$ each of which has 
			different boundary conditions attached to it.\\
			Region 1: $x\in (-\infty,0)$
			\[
			\text{as}\quad x\to-\infty,\quad k_1(x),k_2(x),r(x)\to 0,\quad\quad \text{at }x=0,\quad k_1(x)=k_{11},\quad k_2(x)=k_{12}, \quad r(x)=r_{1}. 
			\]
			Region 2: $x\in (0,a)$
			\[
			\text{at }x=0,\quad k_1(x)=k_{11},\quad k_2(x)=k_{12},\quad r(x)=r_1, \quad\text{at }x=a,\quad k_1(x)=k_{21},\quad k_2(x)=k_{22}, \quad r(x)=r_2.
			\]
			Region 3: $x\in (a,\infty)$
			\[
			\text{at }x=a,\quad k_1(x)=k_{21},\quad k_2(x)=k_{22},\quad r(x)=r_2,\quad \text{as}\quad x\to\infty,\quad k_1(x),k_2(x),r(x)\to 0.
			\]
		\end{enumerate}
		\subsection{Step 6: Energy of the birod}
		We assume small elastic deformations throughout, hence the resulting energy is quadratic in the strain variables. The elastic energy has contributions from the bending of the outer strands 
		eqn. (\ref{eq:kappa}), the extension, bending and twisting of the web eqn. (\ref{eq:microbending}), (\ref{eq:microstrain}) and the stacking energy eqn. (\ref*{eq:int}). 
		\begin{equation}
		E=\int_{-\infty}^{\infty}[\frac{1}{2}EI\kappa^{+2}+\frac{1}{2}EI\kappa^{-2}+\frac{1}{2}\hat{\mathbf{w}}.\mathbf{L}\hat{\mathbf{w}}+\frac{1}{2}\hat{\pmb{\zeta}}.\mathbf{H}\hat{\pmb{\zeta}}+K_e\xi^2+K_ck_3^2]dx.
		\label{Eq:elastic_energy_helrod}
		\end{equation}
		We are especially interested in the interaction energy $\Delta G$ which is the elastic energy of interactions between the two proteins. 
		\eqsp{
			& \Delta G= E^2_{a}-E^1_0-E^1_a,
			\label{Eq:int_energy}
		}
		where $E^2_a$ is the energy of two proteins bound to DNA, one at $x=0$ and other at $x=a$, and $E^1_a$ and $E^1_0$ are the elastic energies corresponding to a single protein binding at $x=a$ and $x=0$, respectively.
		
		\section{Elastic constants} \label{sec:sec5} 
		Our model has $9$ elastic constants $L_1,L_2,L_3,H_1,H_2,H_3,K_c,K_e,EI$. The experimental values for these constants are not known. In order to get some idea about the magnitude of the elastic constants we calculate the extensional modulus, torsional modulus and twist-stretch coupling modulus for a double-stranded DNA within our birod model. The explicit calculation is presented in the appendix. 
		We choose
		\eqsp{
			&K_c=80 \text{pNnm}^2,\quad K_e=600\text{pN}, \quad H_1=15 \text{pN},\quad H_2=10\text{pN},\quad H_3=20\text{pN},\quad L_1=50 \text{pN nm}^{-1},\quad\\
			& L_2=250 \text{pN nm}^{-1},\quad L_3=30 \text{pN nm}^{-1},\quad EI=65 \text{pN nm}^2.
			\label{eqn:elas_const} 
		}
		This choice of elastic constants gives the extensional modulus $S\approx1245$ pN, torsional modulus $C\approx490$ pNnm$^2$ and twist-stretch coupling modulus $g\approx-90$ pNnm which are close to actual 
		values for ds-DNA \cite{singh_acta} measured in experiments. We point out that this choice of elastic constants is not unique, nonetheless we use them to make further calculations.
		
		When we substitute these constants into the governing equations (eqn. ($\ref{eq:12diffeqns}$)) and solve the eigenvalue problem involving $\lambda$, we get the following eigenvalues  
		\eqsp{
			&\lambda_1=-0.68,\lambda_2=-0.42,\lambda_3=-0.36,\lambda_4=0.36,\lambda_5=0.42,\lambda_6=0.68.\quad\quad \text{Units: nm}^{-1}
			\label{eq:exp_sols}
		}
		Other eigenvalues are either very large ($\to \pm\infty$), very small ($\sim 0$) or purely imaginary. Purely imaginary and zero eigenvalues when substituted in $e^{\lambda x}$ give a sinusoidal and a constant 
		function, respectively, which do not decay to zero as $x\to\pm\infty$. As mentioned in section~\ref{sec:step5}, the curvatures $k_1, k_2$ and change in radius $r$ must go to zero at $\pm \infty$. 
		Thus, zero or purely imaginary eigenvalues cannot satisfy our boundary conditions, and are, therefore, not useful. We refer the reader to the appendix for further discussion on the choice of eigenvalues. 
		
		Consider a situation in which two proteins bind DNA, one at $x=0$ and other at $x=a$. In the region $a<x<\infty$ the solution eqn. (\ref{Eq:soln}) consists of only negative eigenvalues. There are three negative eigenvalues $\lambda_{2,3,4}$ and consequently three unknown constants. We have three boundary conditions on $k_1$, $k_2$ and $r$ at $x=a$ to determine those constants. Similarly in the region $-\infty<x<0$, the solution consists of only positive eigenvalues $\lambda_{7,8,9}$, so the constants can again be evaluated from three boundary conditions. We use this scheme to evaluate the strain parameters which we substitute into the expression for the elastic energy functional eqn. (\ref{Eq:elastic_energy_helrod}). Notice that the dominant eigenvalue $\pm0.36$ nm$^{-1}$ corresponds to a decay length of $2.8$ nm ($\approx 10 $ bp) which is what Kim \textit{et al.}\cite{sunscience} report in their experiments. 
		
		\section{Results}
		The experimental evidence for allosteric interactions when two proteins bind to DNA is documented in Kim \textit{et al.}~\cite{sunscience}. Many earlier papers have also described allostery in DNA, but 
		Kim {\it et al.} present exquisite quantitative details which call for a quantitative explanation. 
		
\begin{figure}[H]
			\centering{
				\includegraphics[width=6cm,height=5cm]{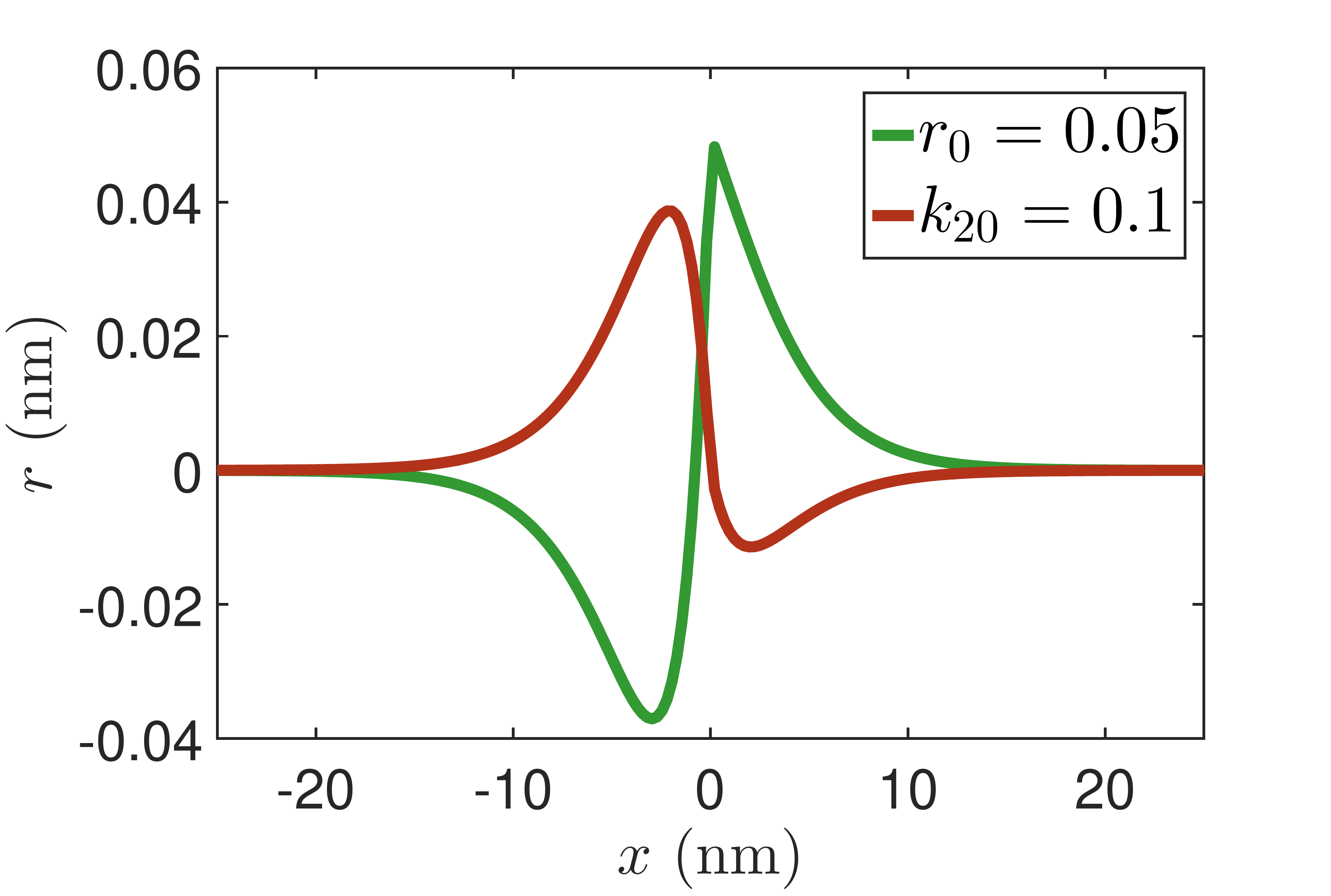}
				\includegraphics[width=6cm,height=5cm]{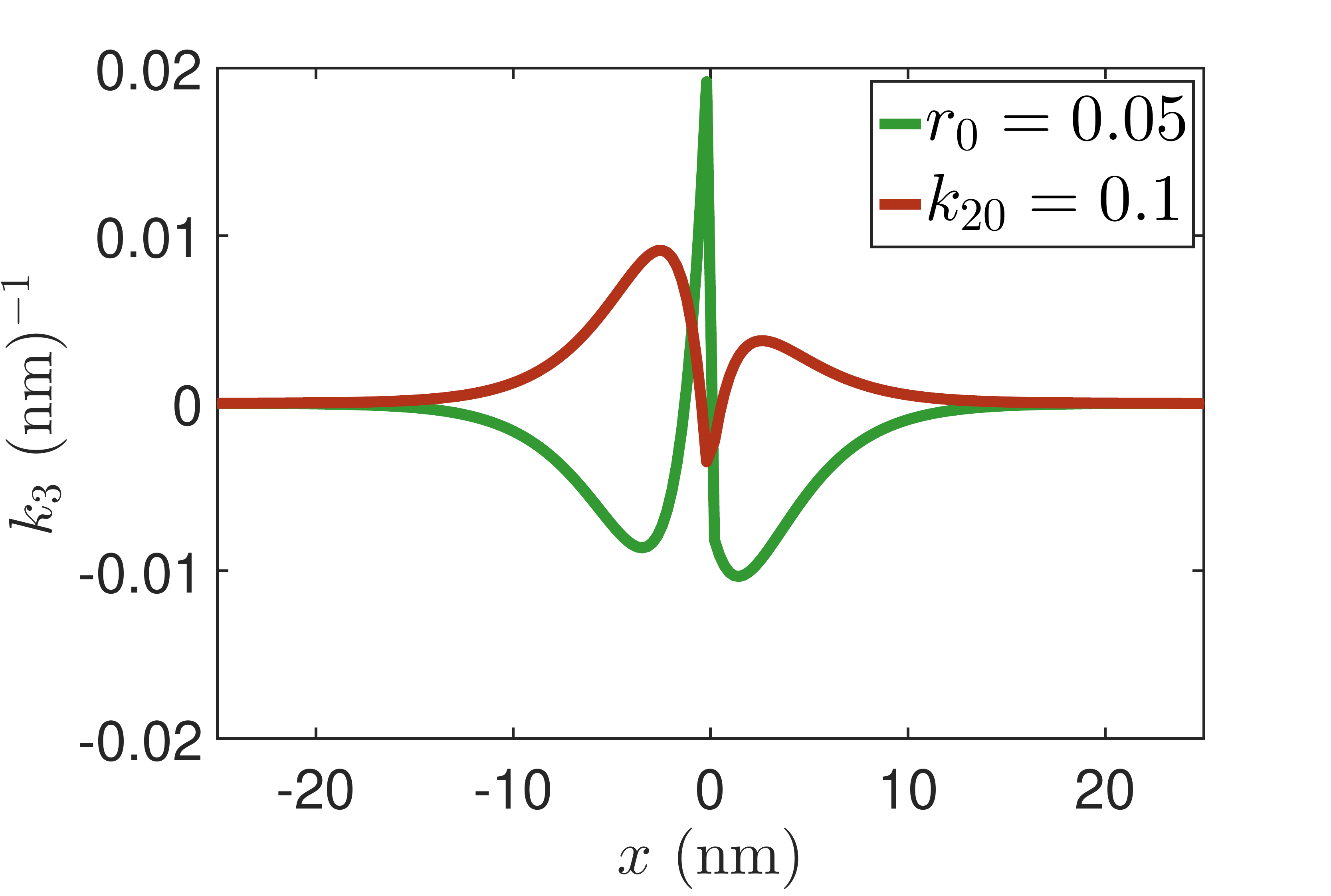}
				\includegraphics[width=6cm,height=5cm]{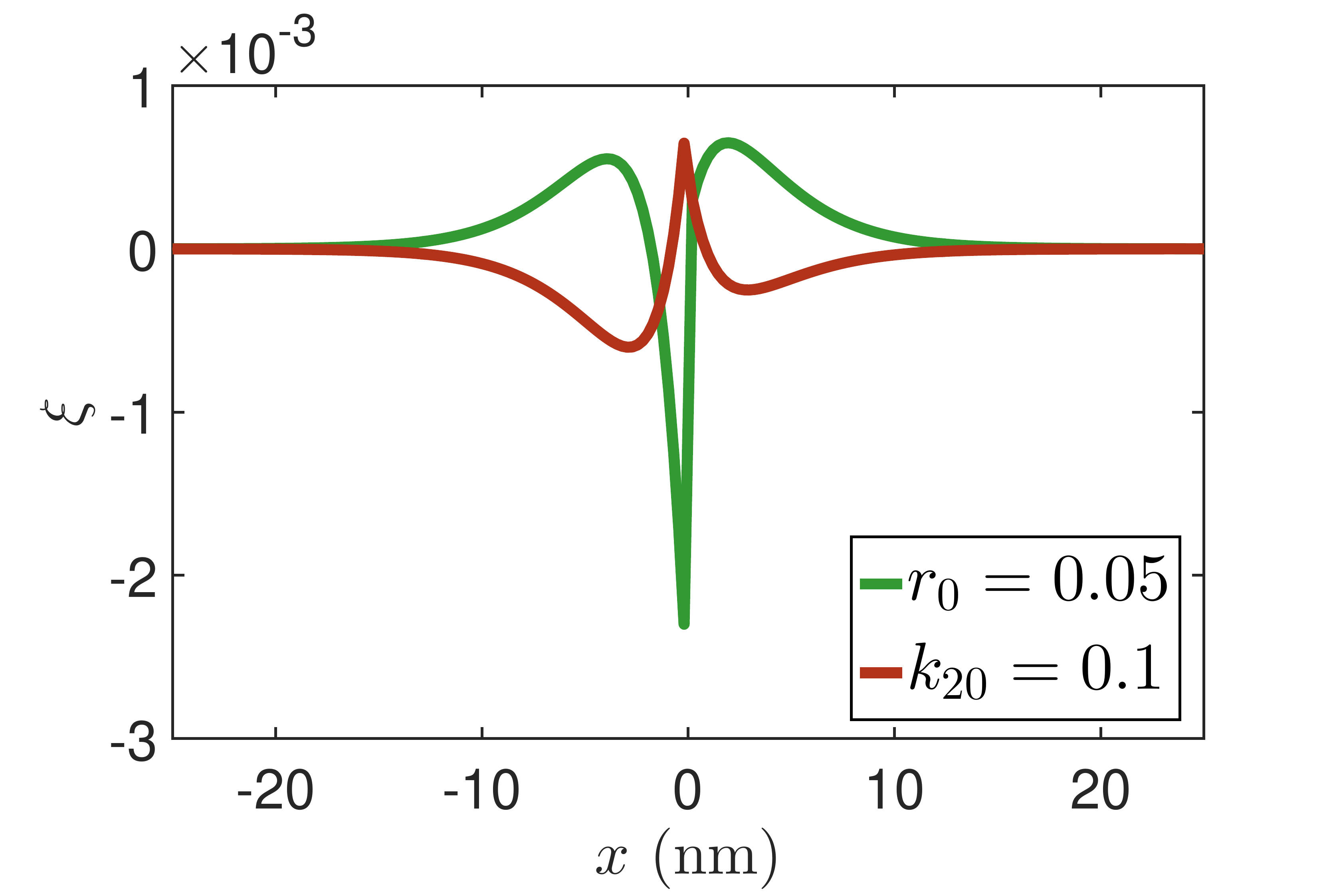}
				\includegraphics[width=6cm,height=5cm]{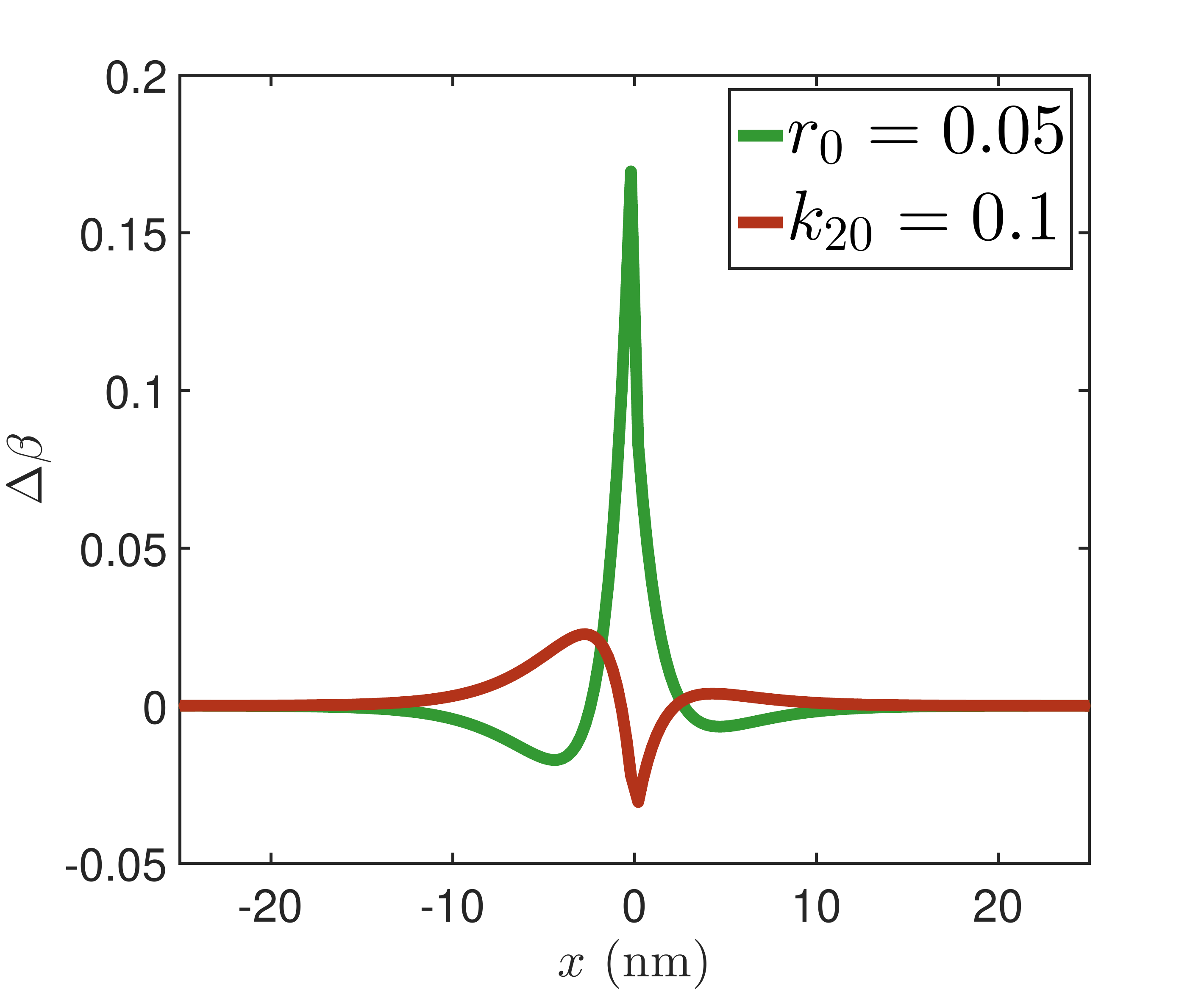}
				\caption{Variation of strain variables for a single protein. We plot the change in radius $r$, twist $k_3$. stretch of centerline $\xi$ and change in phase angle $\Delta \beta=\beta\poss-\beta\negs$ for the double-helix. The red curve correponds to the boundary conditions $k_{10}=r_0=0$ and $k_{20}=0.1\nm^{-1}$ at $x=0$ and the green curve corresponds to $k_{10}=k_{20}=0$ and $r_0=0.05\nm$ at $x=0$. The asymmetry of the double-helix (there is a major and minor groove in DNA) arising from the phase angle $\alpha=2.1$ radian gives the curves a slight asymmetry about the site of protein binding. The curves are exactly symmetric about the site of protein binding if we choose phase angle $\alpha=\pi$ radians (which results in no major and minor groove) as shown in the appendix.}
				\label{fig:1prot1}
			}
		\end{figure}

\begin{figure}[H]
			\centering{
				\includegraphics[scale=0.7]{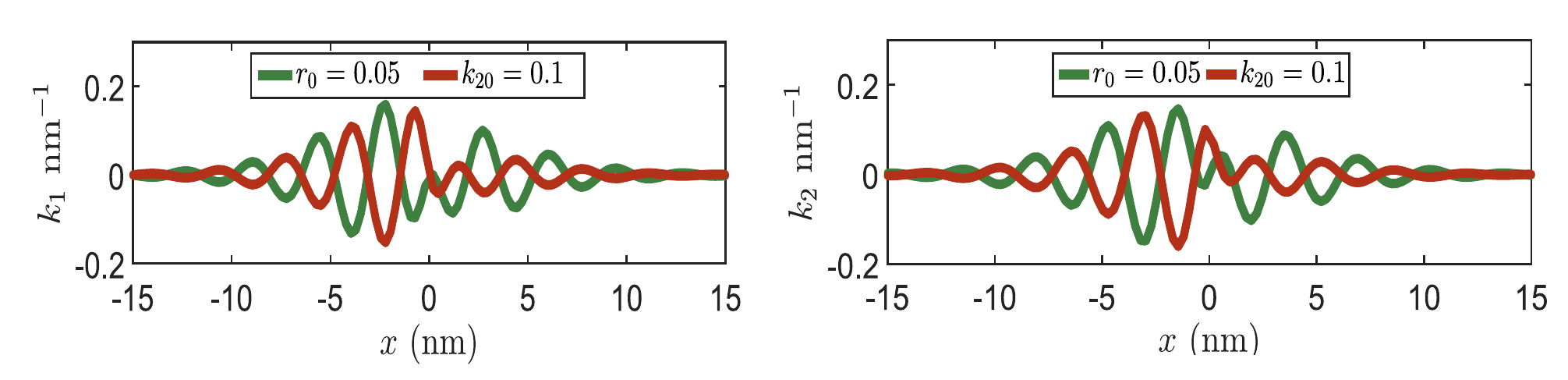}
				\caption{Variation of curvatures $k_1$ and $k_2$ for a single protein. The red curve correponds to the boundary conditions $k_{10}=r_0=0$ and $k_{20}=0.1\nm^{-1}$ at $x=0$ and the green curve corresponds to $k_{10}=k_{20}=0$ and $r_0=0.05\nm$ at $x=0$. We find that the curvature decays exponentially and oscillates with a period $\approx 11$ bp. }
				\label{fig:1prot2}
			}
		\end{figure}

		To unravel the physics behind these allosteric interactions, we begin by examining the case when one protein binds to DNA. As discussed in section \ref{sec:step4}, the strain variables 
		($r,\zeta,\beta^\pm,k_{1,2,3}$) are linear combinations of decaying exponentials. For instance, consider $k_3(x)$ for a protein binding at $x=0$:
		\eqsp{& k_3(x)=p_1\mathbf{v}_{-\lambda}(4)e^{\lambda x} + p_2 \mathbf{v}_{-\mu}(4)e^{\mu x}+p_3\bv_{-\delta}(4)e^{\delta x}\quad\quad x<0,\\
			& k_3(x)=q_1\mathbf{v}_{\lambda}(4)e^{-\lambda x} + q_2 \mathbf{v}_{\mu}(4)e^{-\mu x}+q_3\bv_{\delta}(4)e^{-\delta x}\quad\quad x>0,}
		where $\lambda=0.36\nm^{-1}$, $\mu=0.42\nm^{-1}$, and $\delta=0.68\nm^{-1}$. $\mathbf{v}_{\pm\lambda}$, $\mathbf{v}_{\pm \mu}$, and $\bv_{\pm\delta}$ are the eigenvectors associated with eigenvalues 
		$\pm\lambda$, $\pm\mu$, and $\pm\delta$, respectively. The constants $p_i$ and $q_i$ ($i=1,2,3$) are evaluated using the boundary conditions at $x=0$. It is not difficult to see that the strain variables
		decay to zero as $x\to\pm\infty$. We can replace $k_3$ in the above equation by other strain variables ($r,\xi,\beta^\pm$) and recover similar behavior. We discuss a few characteristics of the variation of the 
		strain parameters as functions of position. The results are plotted in fig.~\ref{fig:1prot1} and fig.~\ref{fig:1prot2}. The strain parameters ($r,k_3,\beta^\pm$) decay exponentially with distance from the site 
		of protein binding. The curvatures exhibit an exponentially decaying sinusoidal character with a period of $11$ bp. This periodic decay of the curvatures manifests itself as sinusoidal variations in the 
		interaction energy. We find that these plots are slightly asymmetric about $x=0$. We attribute this to the structural asymmetry in the right-handed double-helix with phase angle $\alpha=2.1$ radians. If we 
		choose the phase angle $\alpha=\pi$ radians instead, we find that the plots are exactly symmetric about the site of protein binding as shown in the appendix. 
		
\begin{figure}[H]
			\centering{
				\includegraphics[scale=0.4]{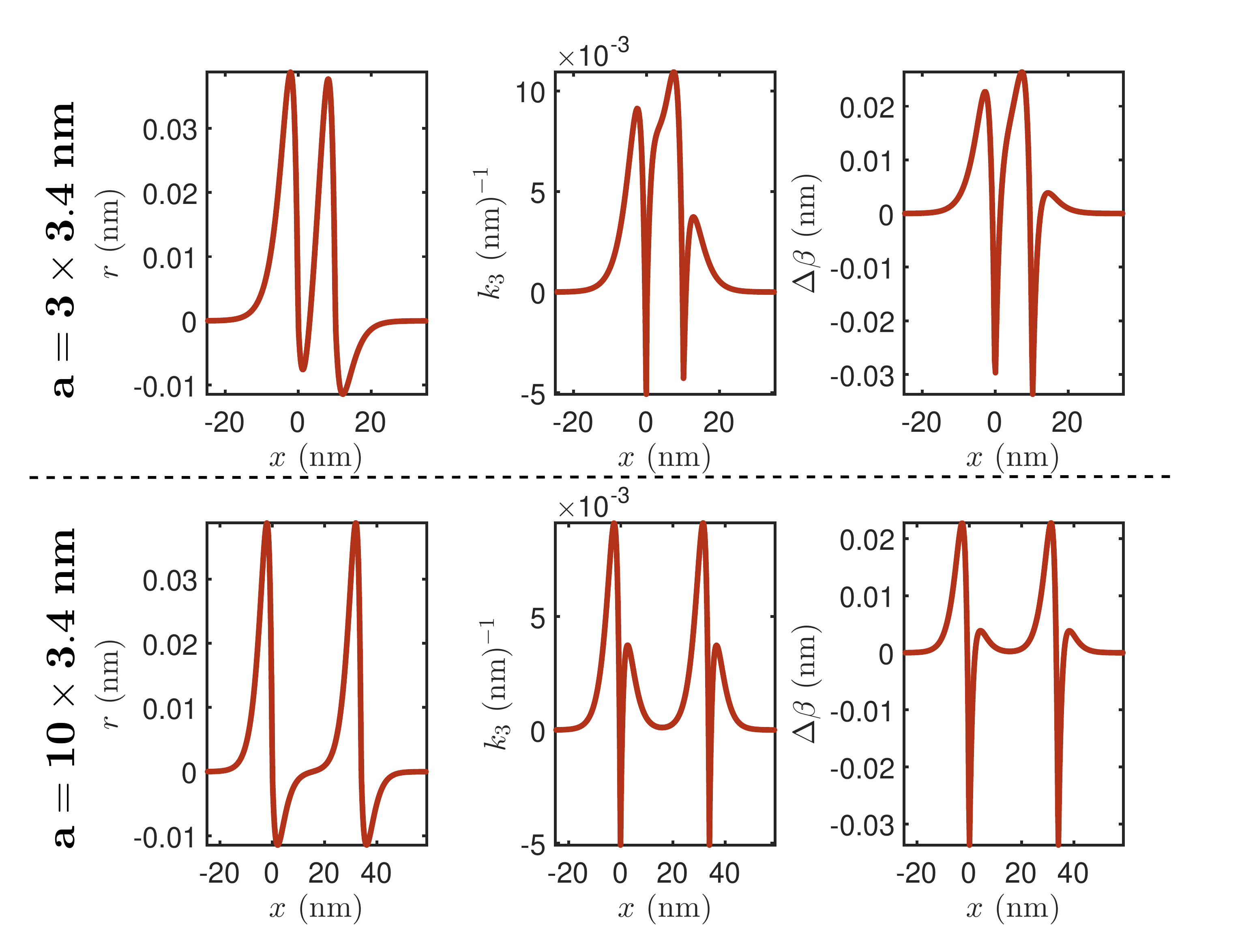}
				\caption{Variation of $r,k_3$ and $\Delta\beta$ for two proteins. Here $a$ is the distance between the sites of protein binding. The strain variables decay exponentially away from the site of protein binding. When the distance between the proteins is large $10\times3.4$ nm, the profile looks like a concatenation of two solutions for a single protein.  }
				\label{fig:2prot1}}
		\end{figure}

\begin{figure}[]
			\centering
			\includegraphics[width=\linewidth,height=6cm]{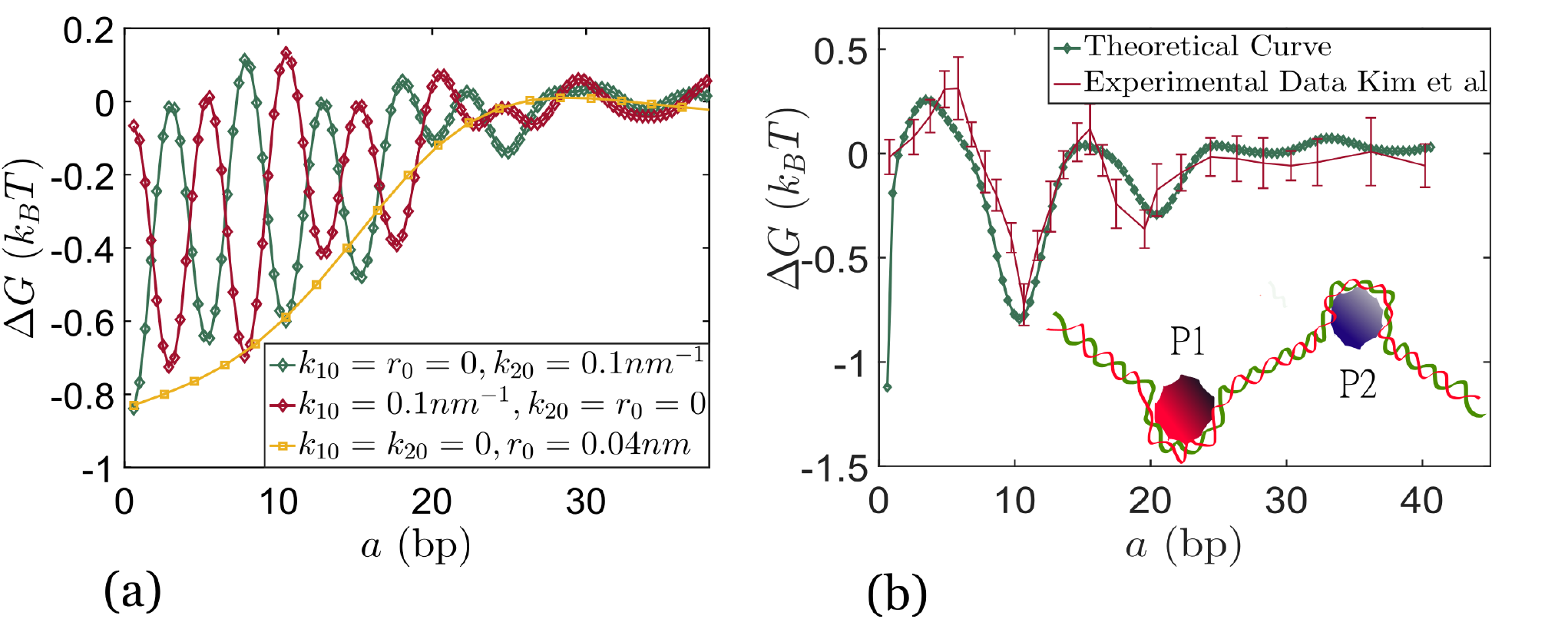}
			\caption{We plot the interaction energy between two proteins eqn. (\ref{Eq:int_energy}). In (a) we plot the behavior of $\Delta G$ for various boundary conditions. If the boundary conditions are specified on the curvatures we get an exponentially decaying profile oscillating with $5-6$ bp $(\approx 11/2$ bp). The oscillatory behavior arises from the periodic geometry of DNA. In (b) the experimental data reproduced for comparison are from Kim \textit{et al}~\cite{sunscience}. We use $k_{11}=k_{21}\approx0.02\nm^{-1}$, $k_{12}=k_{22}=0.05\nm^{-1}$, $r_{1}=-r_{2}=0.02\nm$. The inset in (b) shows a protein DNA complex
				in which the proteins locally bend DNA.}
			\label{fig:deltaG}
		\end{figure}

		We now consider the case when two proteins bind to DNA, one at $x=0$ and the other at $x=a$. We proceed in a similar manner as above and express the strain profiles as linear combinations of exponentials:
		\eqsp{
			\text{Case 1}\quad k_3(x)=&p_1\mathbf{v}_{-\lambda}(4)e^{\lambda x} + p_2 \mathbf{v}_{-\mu}(4)e^{\mu x}+p_3\bv_{-\delta}(4)e^{\delta x}\quad\quad x<0,\\
			\text{Case 2}\quad k_3(x)=&m_1\mathbf{v}_{\lambda}(4)e^{-\lambda x} + m_2 \mathbf{v}_{\mu}(4)e^{-\mu x}+m_3\mathbf{v}_{-\lambda}(4)e^{\lambda x}+\\
			&m_4\mathbf{v}_{-\mu}(4)e^{\mu x}+m_5\bv_{\delta}(4)e^{-\delta x}+m_6\bv_{-\delta}(4)e^{\delta x}\quad \quad0<x<a,\\
			\text{Case 3}\quad k_3(x)=&q_1\mathbf{v}_{\lambda}(4)(3)e^{-\lambda x} + q_2 \mathbf{v}_{\mu}(4)e^{-\mu x}+q_3\bv_{\delta}(4)e^{-\delta x}\quad\quad x>a.}
		The constants $p_i$ and $q_i$ ($i=1,2,3$) are determined by three boundary conditions (on $k_1,k_2$ and $r$) at $x=0$ and $x=a$, respectively. The constants $m_j$, ($j=1,2,3,4,5,6$) are determined by six 
		boundary conditions at $x=0$ and $x=a$. The behavior of the strain variables for two proteins is similar to that for one protein as shown in fig.~\ref{fig:2prot1}.  
		When two proteins are separated by a large distance $a>10\times 3.4$ nm (i.e., more than 10 helical turns of DNA), the strain profile looks like a concatenation of the profiles of two proteins binding 
		separately. Their strain fields do not interact at such distances, thus there is little interaction energy. When the distance decreases, the strain fields of the two proteins overlap, and this is responsible 
		for the interaction energy.
		As discussed in section \ref{sec:sec3}, two defects on a straight ladder fig.~\ref{fig_ladder} interact via an interaction energy that decays exponentially with the distance between them. Now, we focus on the 
		double-helical birod and examine the behavior of different boundary conditions on the interaction energy $\Delta G$ in fig.~\ref{fig:deltaG}. We assume for simplicity that both proteins apply the same boundary 
		conditions on the DNA, the exact numerical values are given in the figure. If we choose the change in radius $r_0=0$ and apply the boundary conditions only on the two curvatures $k_1,k_2$, the interaction 
		energy decays exponentially while varying sinusoidally with a period of $5.5\approx 11/2$ bp. This case corresponds to proteins that bend DNA as shown in the inset of fig.~\ref{deltaG}(b). On the other hand, if the curvatures $k_1,k_2$ are zero 
		while the change the change in radius $r_0$ is non-zero, we get an exponentially decaying profile devoid of any oscillatory character, which is similar to the results for the ladder in section~\ref{sec:sec3}. 
		The exponentially decaying component originates from the elasticity of the web, and the sinusoidal behavior comes from the double-helical structure of DNA. From this exercise we conclude that in order
		to get a sinusoidally varying interaction energy a protein must change the local curvature in the DNA, a mere change in radius of the DNA is not sufficient to give rise to the interaction energy profiles
		observed in experiments.   
		\begin{figure}[H]
			\includegraphics[width=\linewidth,height=7cm]{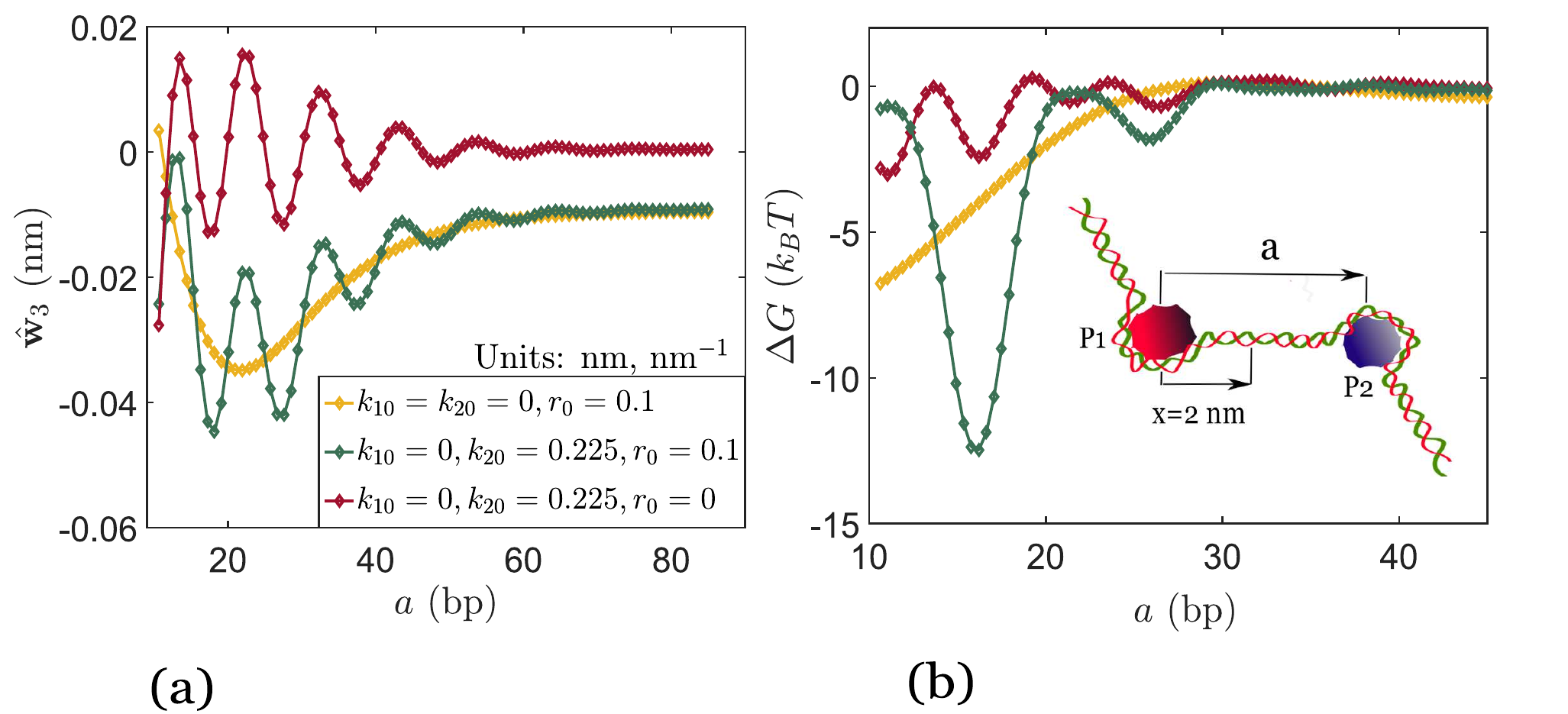}
			\caption{The inset in (b) shows a two protein complex. The boundary conditions are identical for both the proteins $k_{11}=k_{21}=k_{10}$, $k_{12}=k_{22}=k_{20}$, $r_1=r_2=r_0$; the legend in (a) contains the exact numerical values. For (b) the legend is the same as in (a). We examine behavior of $\hat{\bw}_3(x=2$ nm$,a)$ (eqn. \ref{Eq:elastic_energy_helrod}) as a function of distance between the two proteins $a$ for these boundary conditions. The strain variables oscillate with a period of 11 bp. We observe that in case of $r_0=0$, the strain parameter $\hat{\bw}_3(x=2$ nm$,a)$ decays as $e^{-\Gamma a}\psi(\omega a)$ where $\psi(\omega a)$ is a sinusoidal function, hence the combined energy of a two protein complex which is proportional to $(e^{-\Gamma a}\psi(\omega a))^2$ oscillates with a period of 5.5 bp (period of $\sin^{2}x$ is half that of $\sin x$). If $k_{10}=k_{20}=0$ the decay is exponential. If $r_0\neq0$ and $k_{10}$ or $k_{20}\neq 0$,  $\hat{\bw}_3(x=2$ nm$,a) \sim (e^{-\Gamma_1 a}\psi(\omega a) +e^{-\Gamma_2 a})$ and the energy of the two protein complex, which is proportional $(e^{-\Gamma_1 a}\psi(\omega a) +e^{-\Gamma_2 a})^2$, oscillates with a period of 11 bp. The behavior of the other strain variables in eqn (\ref{Eq:elastic_energy_helrod}) is similar. We plot the interaction energy $\Delta G(a)$ in (b) for the boundary conditions indicated in the legend of (a) and use it to verify the period we predict using this argument.}
			\label{fig:deltaG3}
		\end{figure}
		
		In our model the magnitude of the interaction energy increases monotonically with increase in the magnitude of the changes in curvatures or radius caused by the two proteins. Thus, by systematically varying
		the boundary conditions imposed by the proteins we can establish agreement of our theoretical results for $\Delta G$ with the experimental values documented by Kim {\it et al.}\cite{sunscience}. This is done
		in fig.~\ref{fig:deltaG}(b). The values of the curvatures that give the best fit to the experimental data are $k_{11}=k_{21}=0.02$ $nm^{-1}$, $k_{12}=k_{22}=0.05$ $nm^{-1}$ and $r_1=-r_2=0.02\nm$. This choice 
		is, however, not unique and it is coupled with the choice of stiffnesses of the webbing in our birod model. Be that as it may, our exercise above demonstrates that a birod model can capture the dependence 
		of interaction enery on the distance between proteins bound to DNA. Calibration of the model and faithfully connecting it to experiment will require deeper analysis, and perhaps also, computation. 
		
\begin{figure}[H]
			\centering{
				\includegraphics[width=8cm,height=6.5cm]{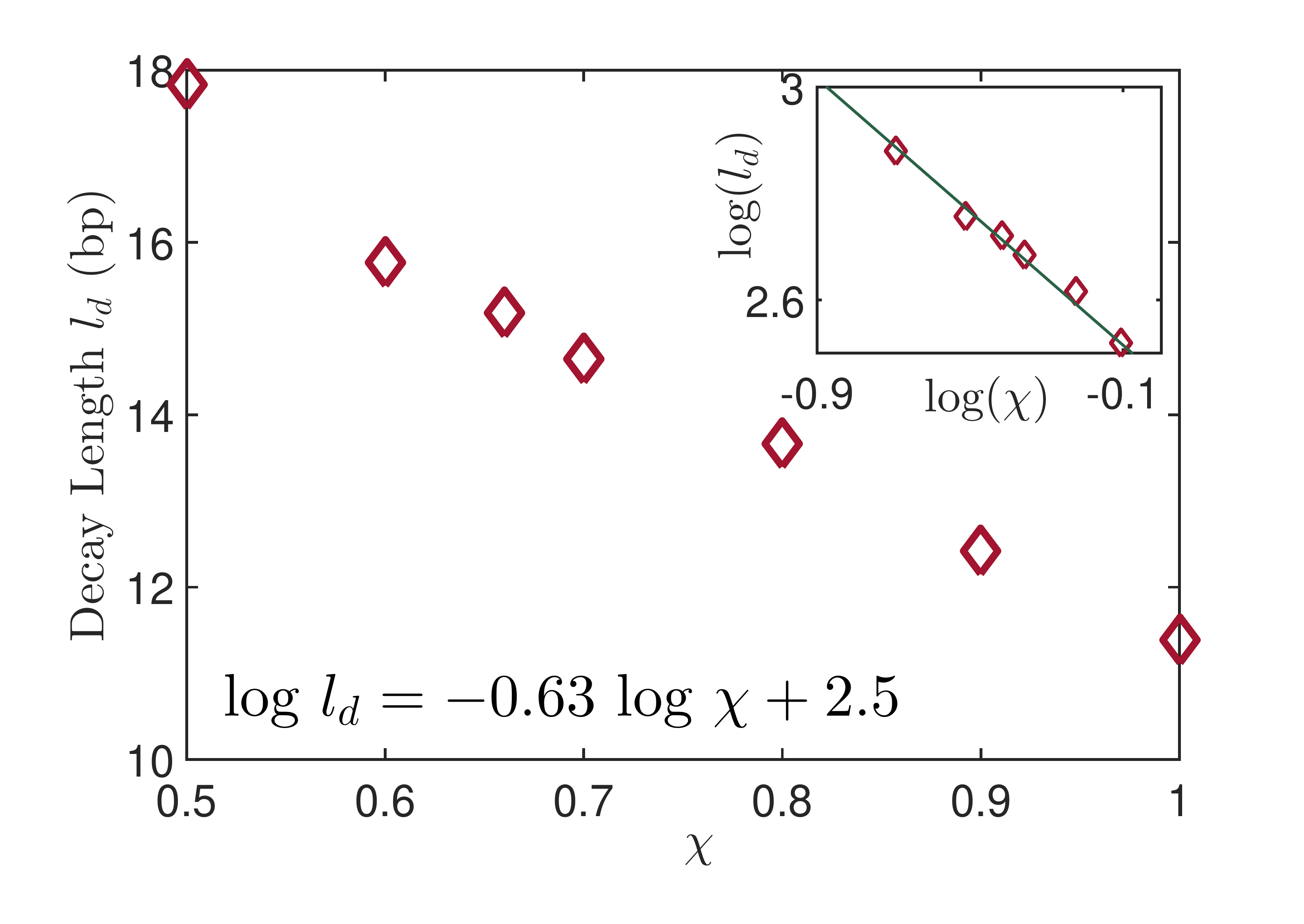}
				\caption{Decay length $l_d$ is defined as the inverse of the eigenvalue with the least non-zero magnitude, for $\chi=1$, $l_d=1/|\lambda|_{min}=1/0.34\approx 9$ bp. $\chi$ is meant to account for the reduction in the elastic constants for AT base-pairs compared to GC base-pairs. The elastic constants for the web are $(\chi K_c, \chi K_e, \chi H_1,\chi H_2,\chi H_3,\chi L_1,\chi L_2,\chi L_3)$, eqn. (\ref{eqn:elas_const}) gives the numerical values for $\chi=1$. We find that the decay length increases with a decrease in elastic constants for the web, thus AT-rich DNA sequences are expected to have higher decay lengths. Qualitative experimental and numerical evidence in support of the above conclusion is documented in \cite{sunscience} and \cite{jpc_simulations}, respectively. The inset shows how we extracted the
					the power law $l_{d} \sim \chi^{-2/3}$.}
				\label{fig:ATGC}
			}
		\end{figure}

		The period the interaction energy in fig. \ref{fig:deltaG}(a) is approximately 5.5 bp while that in fig. \ref{fig:deltaG}(b) is 11 bp as in the experiment. Why? Note that the strain variables in a two protein complex shown in fig. \ref{fig:deltaG3} (b) are a function of both the parameter $x$ and the distance between the two proteins $a$. We fix $x$ ($=2$ nm from protein P$_1$) and focus on the dependence on $a$. We assume that both the proteins apply identical boundary conditions. If the proteins do not cause any change in the radius such that $r_0=0$, then the strain parameters involved in the elastic energy (eqn. (\ref{Eq:elastic_energy_helrod})) $\propto e^{-\Gamma x}\psi(\omega a)$, where $\psi(\omega a)$ is a sinusoidal function oscillating with a period 11 bp, and the elastic energy of the two protein complex $\propto(e^{-\Gamma x}\psi(\omega a))^2$ oscillates with a period 5.5 bp. On the other hand, when the protein causes both a change in radius $r_0$ and a change in curvature $k_{20}$, the strain variables are $\propto (e^{-\Gamma_1 a}\psi(\omega a)+e^{-\Gamma_2 x})$ and the elastic energy of the two protein complex  $\propto (e^{-\Gamma_1 a}\psi(\omega a)+e^{-\Gamma_2 a})^2$ oscillates with a period of 11 bp due to the cross term $e^{-(\Gamma_1+\Gamma_2) a}\psi(\omega a)$. We plot the interaction energy $\Delta G(a)$ between the two proteins constituting the protein complex in fig. \ref{fig:deltaG3}(b) and verify the periods for respective boundary conditions which resolves the apparent discrepancy in the periods in fig.\ref{fig:deltaG} (a) and (b). 
		As a final application of our birod model we examine the sequence dependence of the allosteric interaction energy $\Delta G$. While there is overwhelming qualitative evidence, both experimental \cite{sunscience} and numerical \cite{jpc_simulations}, showing that AT-rich sequences exhibit stronger allosteric interactions compared to GC-rich ones, a theoretical explanation is still lacking. Stronger interactions are 
		associated with longer decay lengths. Using our theory we can find the dependence of the decay length on the elastic constants of the web. Since, AT base-pairs consist of two hydrogen bonds, the corresponding elastic constants for the web are expected to be lower than GC base-pairs which comprise of three hydrogen bonds. In an attempt to simulate such a scenario we replace the elastic constants for the web ($K_c,K_e,L_i,H_i$ $i=1,2,3$) in eqn. (\ref{eqn:elas_const}) with ($\chi K_c,\chi K_e,\chi L_i,\chi H_i$ $i=1,2,3$) while keeping $EI$ fixed, and vary the parameter $\chi$ in the range $0.5\leq\chi\leq 1$. We define a 
		measure of the decay length $l_d$ to be the inverse of the eigenvalue having the least non-zero magnitude, obtained in eqn. (\ref{eq:exp_sols}). For instance, if $\chi=1$, decay length $l_d=1/0.34$ nm$\approx 10$ bp. We plot the variation of $l_d$ with $\chi$ in fig.~\ref{fig:ATGC}. We find that the decay length increases with the decrease in elastic constants of the web. We plot $\log l_d $ versus $\log\chi$ and 
		deduce that $l_d\sim \frac{1}{\chi^{2/3}}$.
		\section{Conclusion}
		Kim {\it et al.}~\cite{sunscience} have presented compelling quantitative evidence for allosteric interactions between two proteins bound to DNA at distant locations. They showed that the interaction energy for two proteins separated by distance $a$ on DNA is a decaying exponential oscillating with period of 11 bp. Various attempts to numerically simulate the allosteric interactions have been made \cite{jpc_simulations,nacid_simulations} and have associated the oscillating interaction energy to the major groove width in the double-helical structure of DNA. We approach the problem from a purely mechanical standpoint. We conjecture that the local deformation field in DNA caused by a bound protein is similar to that produced by a defect in an elastic solid. We begin by computing the interaction energy for two defects on a ladder 
		and find that it decays exponentially with the distance between them. We, then, proceed to replicate the same calculation for DNA by modelling it as a double-helical birod~\cite{maddocks}. We assume that 
		the outer phosphate backbones represented by $\pm$ strands to be inextensible and 
		unshearable while the base-pairs are capable of elastic extension, shear, bending and, twisting. We assume a general form of displacement for these strands (eqn. \ref{eq_r_def}) which we use to calculate 
		the micro-displacement and micro-rotation for the base-pairs. We, then, use these expressions to solve the governing equations for our birod. A crucial factor in our treatment is the boundary conditions. 
		We follow Kwiecinski \textit{et al.}~\cite{curv_sensing_prot_goriely}, Kim \textit{et al.}~\cite{sunscience} and Liang and Purohit~\cite{liang_membrance} and impose boundary conditions on the curvatures 
		and the radius of the DNA double-helix. The question, ``what kind of boundary conditions a protein could possibly apply", is not yet comprehensively addressed in the literature and is not the central issue 
		of this study either. Rather our message is that after solving the governing equations and plugging in boundary conditions, we recover the exponentially decaying profile that oscillates with a period of 
		11 bp. We end by examining the sequence dependence of allosteric interactions and show that AT-rich sequences exhibit stronger interactions than GC-rich sequences.  
		
		Even though our birod model does surprisingly well by capturing the dependence of interaction energy on distance there are many important caveats that we must point out. First, we do not expect our birod model
		to be accurate near the site of protein binding. The deformations near the binding site could be large enough that a linear elastic theory may not be applicable. Our assumptions that the outer strands are 
		inextensible and the web is elastic could also break down in the vicinity of the binding site. Second, we have little knowledge of the elastic constants of the web. We have assumed some stiffness parameters
		for the web that gave the right experimentally verified moduli for the DNA, but there could have been another set of parameters that would have given similar results. One may have to appeal to molecular 
		simulations \cite{maddocks_rigidbp,olson1,olsen2,olsen3,olson4} to get these parameters. Third, the boundary conditions applied by the proteins on the DNA are not clear. One may have to look for guidance 
		from molecular simulations or protein-DNA co-crystal structures to get a clearer picture. Finally, we have not accounted for fluctuations or entropic interactions in our model. This is partly justifiable 
		because the length of DNA between two protein binding sites for which significant allosteric interactions are observed is often much smaller than the persistence length of the DNA. However, a rigorous 
		calculation should be done to verify this assumption. In spite of these shortcomings, our model could provide a starting point for analyzing allosteric interactions in DNA within the broad framework of
		configurational forces in elastic solids.\\  
		\vspace*{10mm}\\
		\textit{We acknowledge insightful discussion with Yujie Sun who is one of the authors in Kim \textit{et al.} \cite{sunscience}.}\\
		%%%%%%%%%% Insert bibliography here %%%%%%%%%%%%%%

		\appendix
		\section{Appendix}
		\subsection{Kinematics of the $-$ strand} 
		In the main text we gave detailed derivations for the strains, curvatures, etc., for the $+$ strand in our birod. 
		We now shift our attention to the complimentary $-$ strand. The reference configuration of this strand is denoted by position vector $\brn_0$.
		\begin{equation}
		\brn_0=b(\cos (\omega x+\alpha)\be_1+\sin(\omega x+\alpha)\be_2)+x\be_3.
		\end{equation}
		Along the same lines as the $+$ strand, we conceive the deformed configuration to be a helix wrapped around a curved axis defined by curvatures $k_1,k_2$ and $k_3$ along the directors 
		$\bd_1,\bd_2$ and $\bd_3$, respectively. 
		\begin{equation}
		\begin{split}
		&\brn(x)=(b+r^-)(\cos (\omega x+\alpha+\beta\negs)\bd_1+\sin (\omega x+\alpha +\beta\negs)\bd_2)+\int_{0}^{x}(1+b\xi)\bd_3dx.
		\end{split}
		\end{equation}
		We use the same apparatus \textit{mutatis mutandis} described for the $+$ strand to calculate various quantities of interest. The results are:
		\begin{equation}
		\bRn=[\nn\quad\bn\quad\tn]=\mathbf{Z}\bRn_0(\mathbf{1}+\thetan).
		\end{equation} 
		where $\thetan$ is a skew symmetric tensor.
		\begin{equation}
		\begin{split}
		&\thetan=\begin{bmatrix}
		0 &-\theta^-_3&\theta_2\\\theta^-_3&0&-\theta^-_1\\-\theta^-_2&\theta^-_1&0
		\end{bmatrix},\\
		\text{where}\quad& \theta^-_1=(r\negs\omega+b(\beta^-_x+k_3)),\quad \theta^-_2=-r\negs_x\cos k+\beta^- \sin k, \quad \theta^-_3=\frac{g\negs}{\omega \sin k}-\frac{(r\negs_x\cos k-\beta^- \sin k)\cos k}{\omega \sin k}.\\
		\end{split}
		\end{equation}
		We compute curvature $\kappa\negs$ as follows,
		\begin{equation}
		\begin{split}
		\Omega\negs=&(\tn_x.\tn_x)^{1/2}=\omega\sin k-(r\negs_{xx}+\xi)\cos k+(\beta\negs_{x}+k_3)\sin k,\\
		\kappa\negs=&\Omega\negs-\omega \sin k=-(r\negs_{xx}+\xi)\cos k+(\beta\negs_{x}+k_3)\sin k.
		\end{split}
		\end{equation}
		We obtain the moment $\bimn$ as follows,
		\eqsp{&\bimn=EI\kappa\negs(\cos k \cosabt \bfv_1+\cos k \sinabt\bfv_2+\sin k \bfv_3),}
		where $\bfv_1,\bfv_2,\bfv_3$ are given by eqn. \ref{eq:Q}.

		%\begin{figure}[H]
		%	\centering{
		%		
		%		\caption{Variation of curvatures $k_1,k_2$ for $\alpha=\pi$. }
		%		\label{fig:1prot_curv_pi}
		%	}
		%\end{figure}
		
		\subsection{Evaluation of material properties of the web}
		In this section, we consider a deformation of the double-helical structure induced by a stretching force $F$ and torque $T$ on one end. We assume that the helix retains its helical configuration, but with changed geometrical parameters. Thus, $r$, $\beta$ and $e$ are independent of $x$. Our goal is to compute the strains and curvatures, then evaluate the energy, and then identify the stretch
		modulus, twist modulus and twist-stretch coupling modulus of the double-helical structure from this energy expression. The computation of strains, curvatures, etc., of the helix proceeds as in the main 
		text. 
		\begin{equation}
		\begin{split}
		&\brp=(a+r)(\cosr(1+\beta)\be_1+\sinr(1+\beta)\be_2)+x(1+e),\\
		&\brp=-(a+r)(\cosr(1+\beta)\be_1+\sinr(1+\beta)\be_2)+x(1+e),\\
		\end{split}
		\end{equation} 
		We assume $r,\beta,e\sim \oeps$, hence
		\begin{equation}
		\begin{split}
		\brp&=(a+r)\er+a\omega \beta x\ethe+x(1+e)\be_3,\\
		\brp_x&=(a+r)\omega\ethe+a\omega\beta\ethe-a\omega^2,\\
		&\beta x\er+(1+(ex)_x)\be_3,\\
		&=-a\omega^2\beta x\er+\omega(a+r+a\beta)\ethe+(1+(ex)_x)\be_3.
		\end{split}
		\end{equation}
		The inextensibility condition gives,
		\begin{equation}
		\begin{split}
		&|\brp_x|=|\brp_{0x}|,\\
		&(ex)_x+\omega^2a(r+\beta)=0,
		\end{split}
		\label{Eq:appendix_inext}
		\end{equation}
		$\tpz$, $\np_0$ and $\bpz$ are the tangent, normal and binormal to the $+$ strand in the reference configuration. We calculate tangent $\tp$ to the deformed configuration. 
		\begin{equation}
		\begin{split}
		\tp=&-\sin k \beta x\er+(\sin k+\omega r\cos k+\beta\sin k )\\
		&\ethe+(\cos k-\omega \sin k(r+a\beta))\be_3,\\
		=&\tpz+\omega \beta x \sin k~\pmb{n}_0\poss+(\omega r+\beta \tan k)\bpz,
		\end{split}
		\end{equation}
		Next, we calculate the curvature $\kappa \poss$.
		\begin{equation}
		\begin{split}
		\tp_x=&-(\omega \sin k 2\omega \beta \sin k+\omega^2 r \cos k)\er\\
		&-\omega^2\sin k \beta x \ethe.\\
		K^2=&\omega \sin k +2\omega \beta \sin k+\omega^2 r\cos k.\\
		\kappa\poss=&K-\omega\sin k=2\omega \beta \sin k+\omega ^2 r \cos k.\\
		\end{split}
		\end{equation}
		We go on to calculate the normal in the deformed configuration $\np$. 
		\begin{equation}
		\begin{split}
		\np=&-\er-\omega \beta x \ethe,\\
		=&\np_0-\omega \beta x \sin k \tpz+\omega \beta x \cos k \bpz.
		\end{split}
		\end{equation}
		We are now in a position to calculate the deformed Frenet-Serret frame $\bRp$.
		\begin{equation}
		\bRp=[\np\quad\bp\quad\tp]=\bRp_0(\mathbf{1}+\thetap).
		\end{equation} 
		where $\thetap$ is a skew symmetric tensor.
		\begin{equation}
		\begin{split}
		&\thetap=\begin{bmatrix}
		0 &-\theta^+_3&\theta^+_2\\\theta^+_3&0&-\theta^+_1\\-\theta^+_2&\theta^+_1&0
		\end{bmatrix},\\
		\text{where}\quad\quad& \theta^+_1=\omega r+\beta \tan k, \quad  \theta^+_2=\omega \beta x \sin k,\quad \theta^+_3=\omega \beta x \cos k.
		\end{split}
		\end{equation}
		For the negative strand we follow the same procedure.
		\begin{equation}
		\begin{split}
		&\bRn=[\nn\quad\bn\quad\tn]=\bRn_0(\mathbf{1}+\thetan),\\
		&\thetan=\thetap,\\
		&\kappa\negs=\kappa\poss.
		\end{split}
		\end{equation}
		After performing all the calculations
		\begin{equation}
		\begin{split}
		E=\int_{0}^{L}(EI(2\omega \beta& \sin k+\omega^2 r\cos k)^2+\frac{1}{2}H_1\omega^2(r+a\beta)^2+\frac{1}{2}L_1 r^2)-M\theta -F\Delta x,\\
		&\Delta x=eL,\quad\quad\theta=\beta L.\\
		\end{split}
		\end{equation}
		We substitute $r=-\frac{e}{\omega ^2 a}-a\beta$ from eqn. (\ref{Eq:appendix_inext}) and compute the elastic constants as follows. 
		\begin{equation}
		\begin{split}
		&\frac{\partial E}{\partial \beta}=0,\quad\quad \frac{\partial E}{\partial e}=0.\\
		&S=\frac{\partial^2 E}{\partial e^2},\quad g=\frac{\partial^2 E}{\partial e\partial \beta},\quad C=\frac{\partial^2 E}{\partial  \beta^2}.
		\end{split}
		\end{equation}
		Then, by trial and error we pick values of $L_{1}, L_{2}, L_{3}, H_{1}, H_{2}, H_{3}, K_{c}, K_{e}, EI$ to match the $S, g, C$ known from
		experiments. Our choice of the material parameters $L_{1}, H_{2}, K_{c}$, etc., is not unique.  
		
		\subsection{Choice of eigenvalues obtained in section \ref{sec:sec5}}
		In section \ref{sec:sec5}, we solve the governing differential equation eqn. $\ref{eq:12diffeqns}$ by substituting $y(x)=y_0e^{-\lambda x}$ where $y=(r,f,\xi,k_3,\beta^\pm,n^c_i,n_i)\quad$ $i=1,2,3$. We look for the values of $\lambda$ corresponding to a non-trivial solution of the governing equations. For this we need to solve the eigenvalue problem $\mathcal{A}(\lambda)\mathbf{v}_0=0$, where $\mathcal{A}$ is a function of $\lambda$ and elastic constants (eqn. \ref{eqn:elas_const}) and $\mathbf{v}_0=[r_0,f_0,\xi_0,k_{30}, \beta\poss_0,\beta\negs_0, n^c_{i0},n_{i0}]^T\quad$   $i=1,2,3$. We set $\det \mathcal{A}(\lambda)=0$ and get following solutions for $\lambda$.
		\eqsp{
			&x_1=-1.5\times 10^4(1+i),\quad x_2=-1.5\times 10^4(-1+i),\quad x_3=-4\times10^3,\quad x_4=1.2\times10^3(-1-3.2i),\\
			&x_5=1.2\times10^3(-1+3.2i), \quad x_6=-0.68,\quad x_7=-0.42,\quad x_8=-0.36,\quad x_9=-5.2\times10^{-10},\\
			&x_{10}=-1.9i,\quad x_{11}=1.9i,\quad x_{12}=-3.8i,\quad x_{13}=3.8i,\quad x_{14}=-6.2i,\quad x_{15}=6.2i,\\
			& x_{16}=5.2\times10^{-10},\quad x_{17}=0.36,\quad x_{18}=0.42,\quad x_{19}=0.68,\quad x_{20}=2.3\times 10^3(1.4-i),\\
			&x_{21}=2.3\times 10^3(1.4+i),\quad x_{22}=1.5\times 10^4(1-i),\quad x_{23}=1.5\times 10^4(1+i).}
		Among these 23 eigenvalues we neglect the eigenvalues $x_{1,2,3,4,5,20,21,22,23}$ whose magnitude is $>10^3$ because the corresponding decay length is tiny which leads to large numerical errors given that we need to compute third derivatives. Then, there are small eigenvalues $x_{9,16}$ whose magnitude is close to zero $(<10^{-3})$ and purely imaginary eigenvalues $x_{10,11,12,13,14,15}$ which when substituted in $e^{-\lambda x}$ result in a constant or a sinusoidal function, respectively, that do not decay to $0$  as $x\to\pm\infty$. Hence, we must neglect these too. This leaves us with $x_{6,7,8,17,18,19}$, which are used in section \ref{sec:sec5}.  
		
		\subsection{Results for $\alpha=\pi$ radians}
		\begin{figure}[H]
			\centering{
				\includegraphics[width=6cm,height=3.5cm]{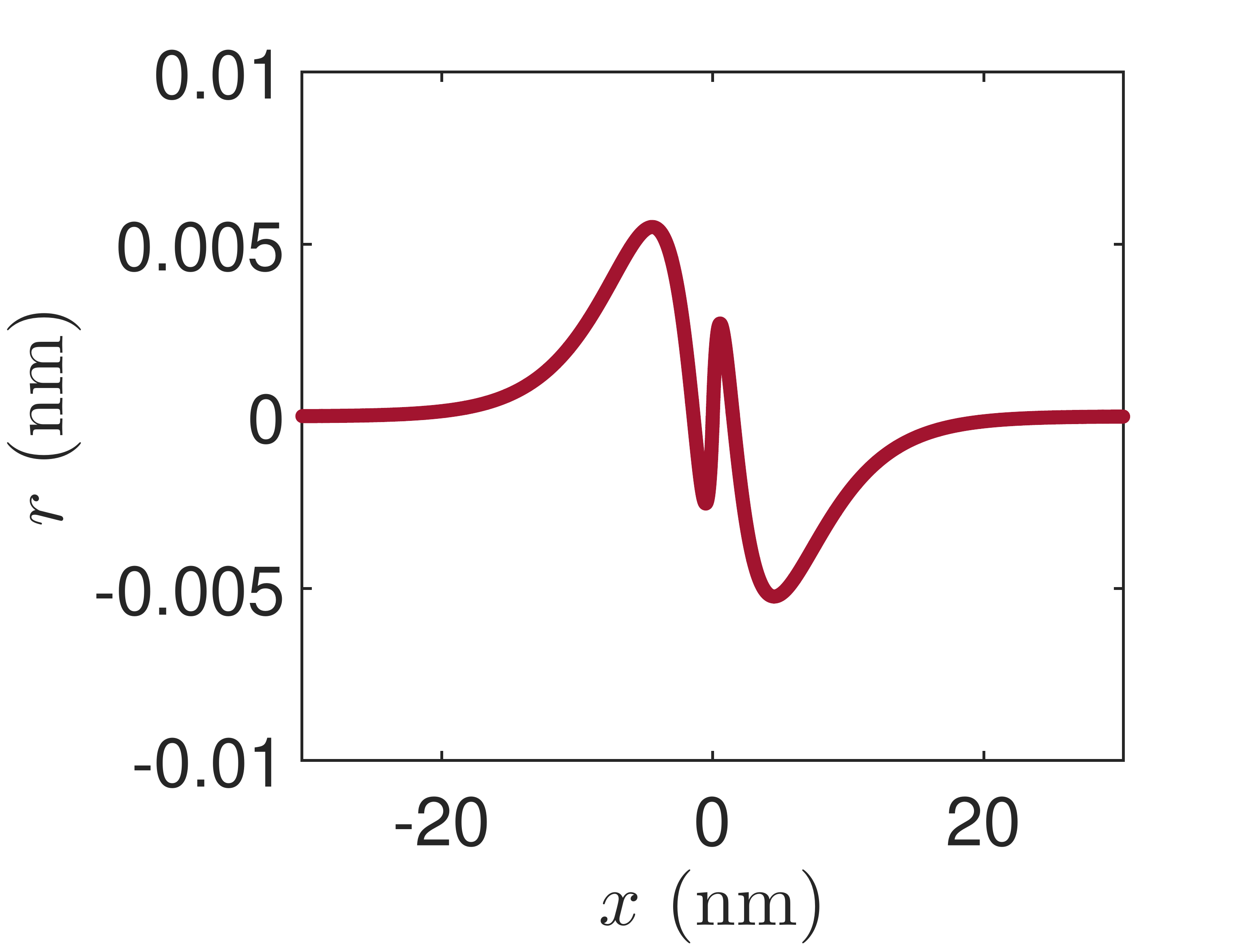}
				\includegraphics[width=6cm,height=3.5cm]{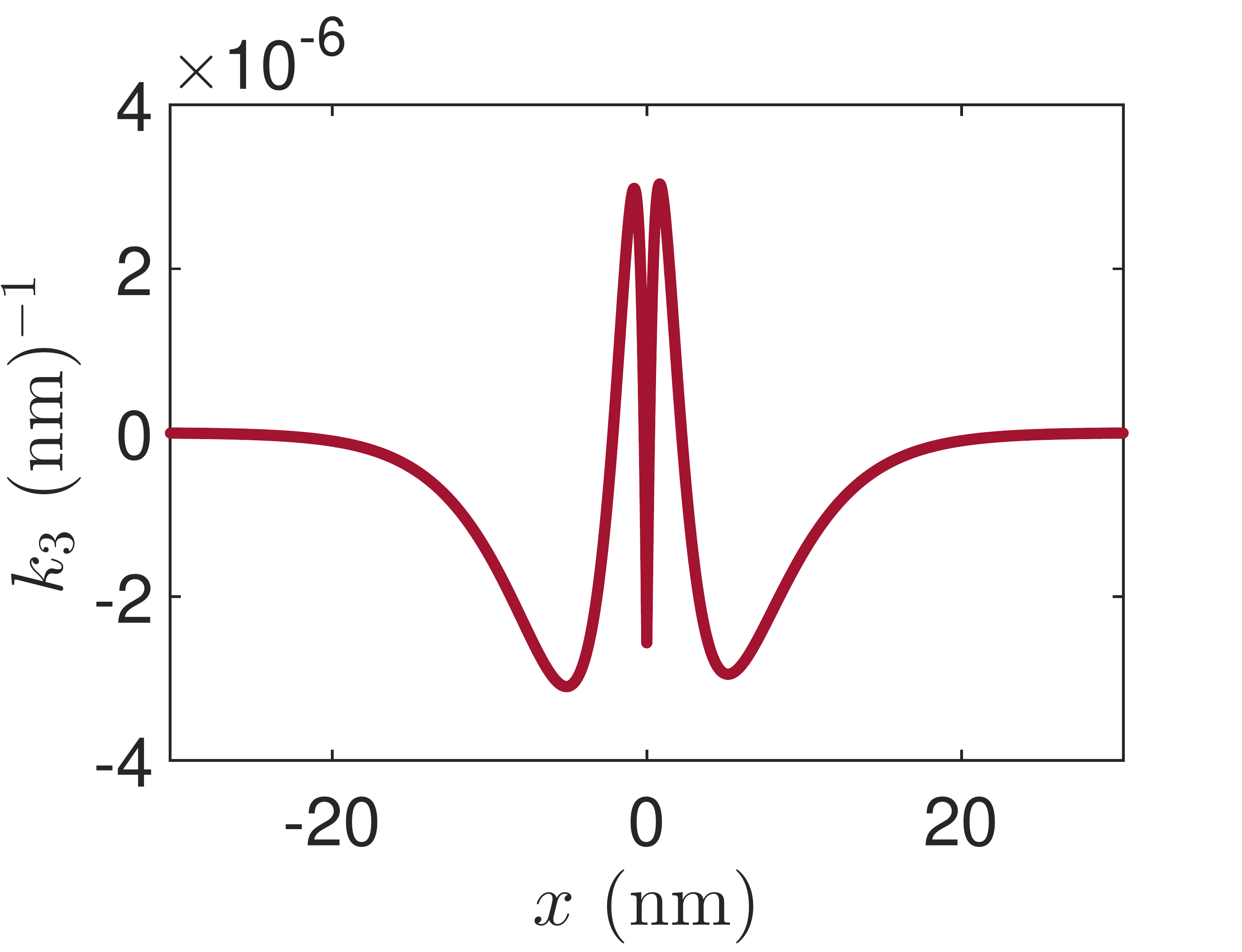}
				\includegraphics[width=6cm,height=3.5cm]{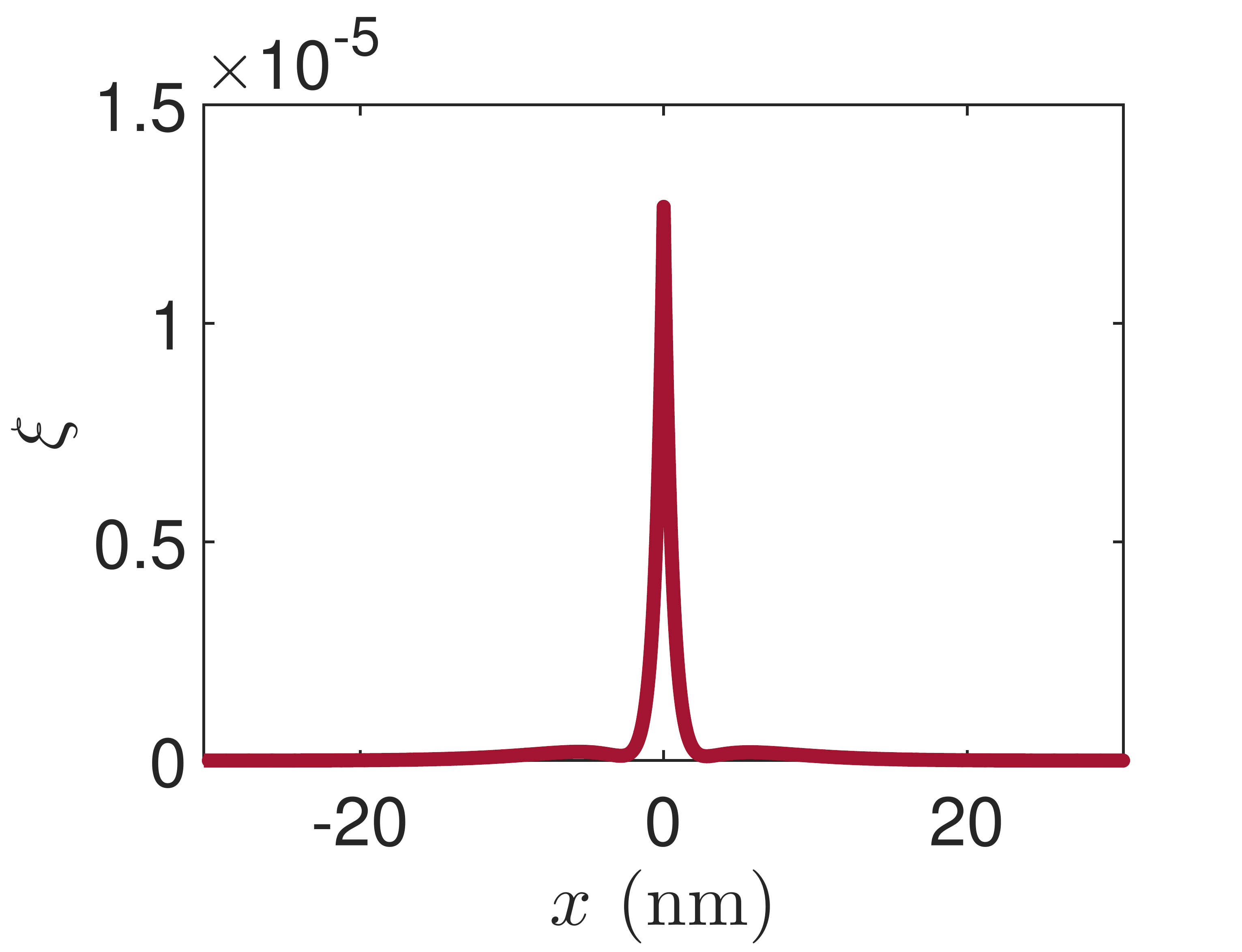}
				\includegraphics[width=6cm,height=3.5cm]{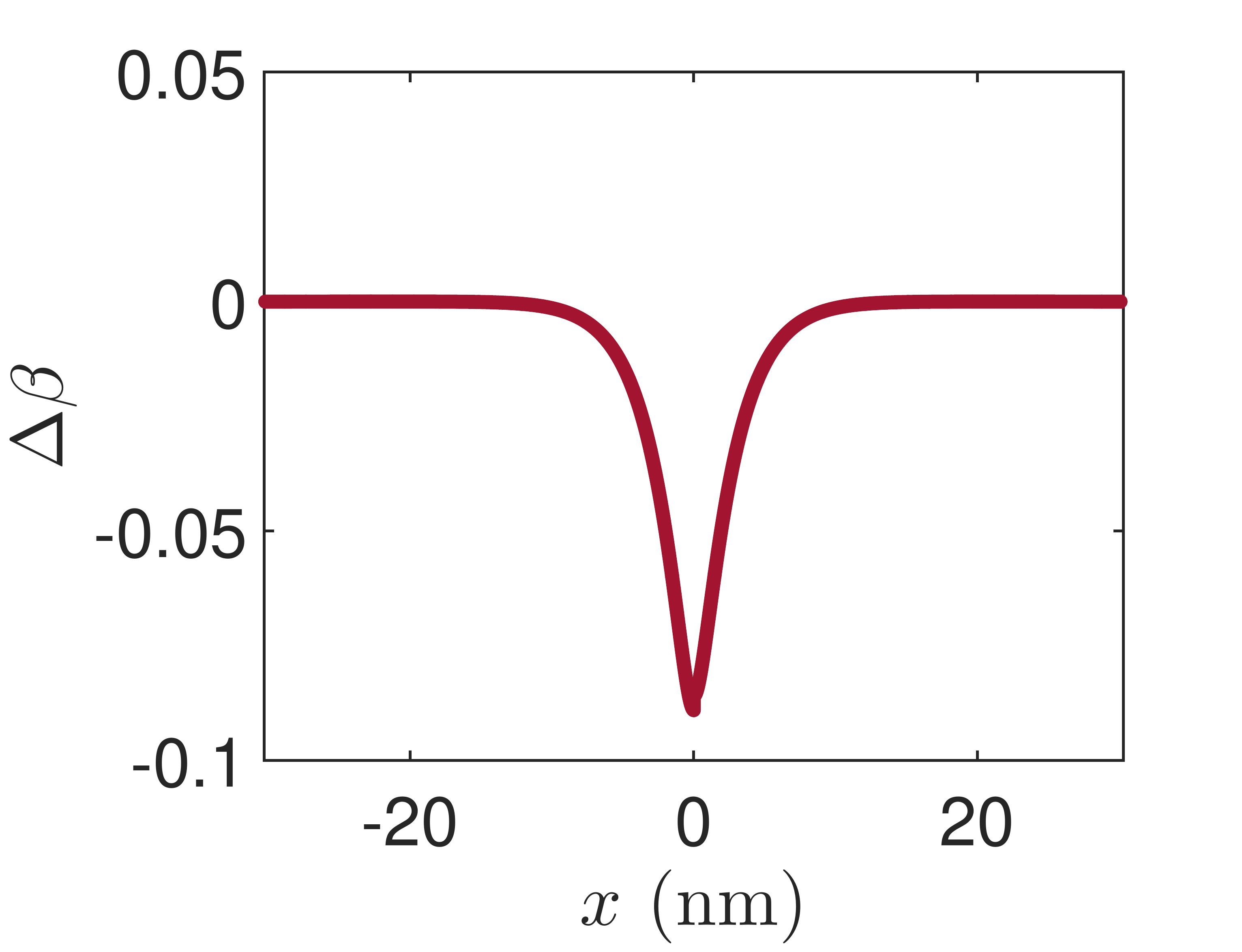}
				\includegraphics[width=8.5cm,height=3.5cm]{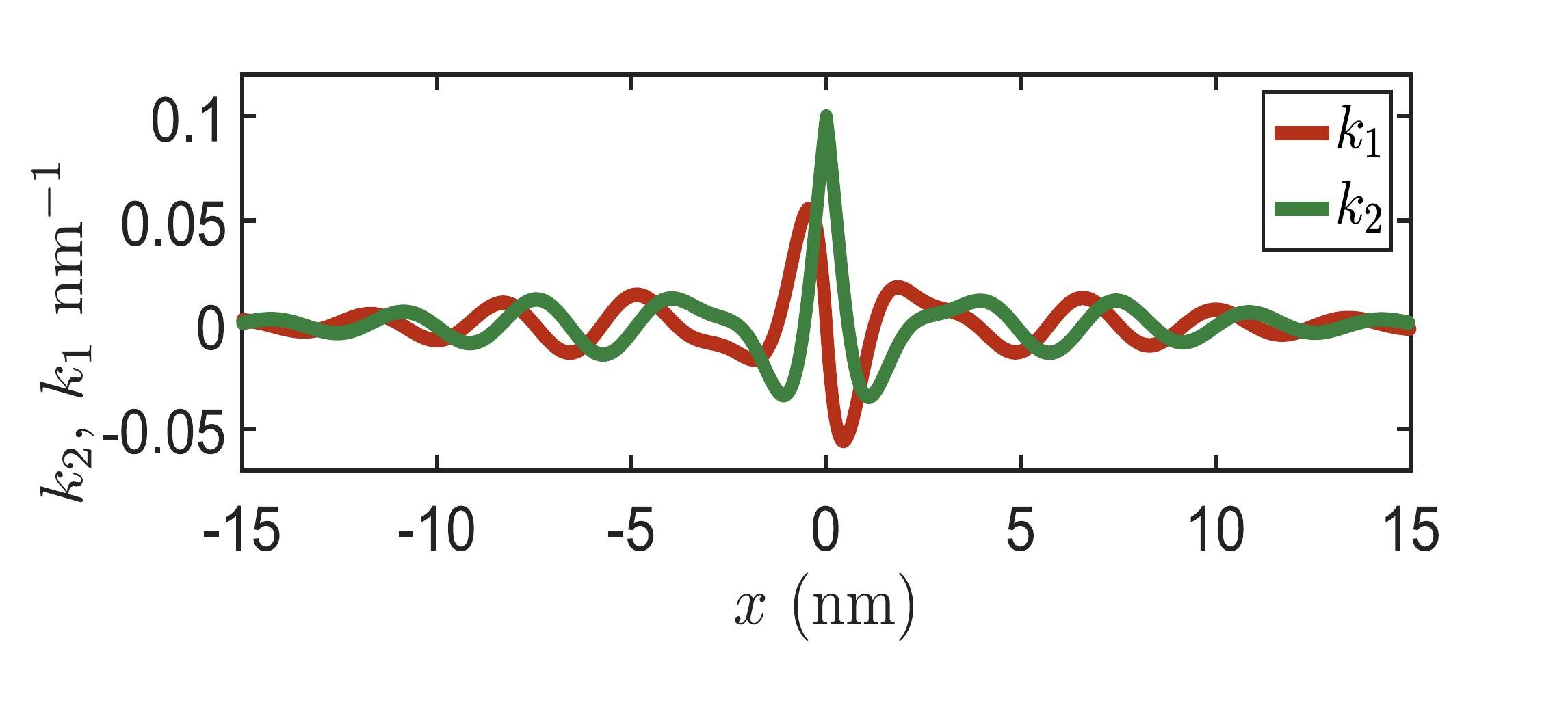}
				\caption{Variation of strain variables for $\alpha=\pi$ radians. Notice that the curves are symmetric about the site of protein binding. As mentioned in section 6, the curves are not symmetric if we choose $\alpha=2.1$ radians.}
				\label{fig:1prot_pi}
			}
		\end{figure}
		
	\end{document}